\documentclass[preprint,12pt]{elsarticle}

\usepackage{amssymb}
\usepackage{amsmath}
\usepackage{caption}
\usepackage{booktabs}
\usepackage{longtable}
\usepackage{multirow}
\usepackage{array}
\usepackage{subfigure}
\usepackage{subcaption}
\usepackage{float}
\usepackage{hyperref}

\journal{Neural Networks}

\begin{document}

\begin{frontmatter}



\title{Meta-Learning-Based Delayless Subband Adaptive Filter using Complex Self-Attention for Active Noise Control}


\author[1]{Pengxing Feng} 
\author[1]{Hing Cheung So} 
\affiliation[1]{organization={Department of Electrical Engineering, City University of Hong Kong},
            city={Hong Kong SAR},
            country={China}}
\begin{abstract}
Active noise control typically employs adaptive filtering to generate secondary noise, where the least mean square algorithm is the most widely used. However, traditional updating rules are linear and exhibit limited effectiveness in addressing nonlinear environments and nonstationary noise. To tackle this challenge, we reformulate the active noise control problem as a meta-learning problem and propose a meta-learning-based delayless subband adaptive filter with deep neural networks. The core idea is to utilize a neural network as an adaptive algorithm that can adapt to different environments and types of noise. The neural network will train under noisy observations, implying that it recognizes the optimized updating rule without true labels. A single-headed attention recurrent neural network is devised with learnable feature embedding to update the adaptive filter weight efficiently, enabling accurate computation of the secondary source to attenuate the unwanted primary noise. In order to relax the time constraint on updating the adaptive filter weights, the delayless subband architecture is employed, which will allow the system to be updated less frequently as the downsampling factor increases. In addition, the delayless subband architecture does not introduce additional time delays in active noise control systems. A skip updating strategy is introduced to decrease the updating frequency further so that machines with limited resources have more possibility to board our meta-learning-based model. Extensive multi-condition training ensures generalization and robustness against various types of noise and environments. Simulation results demonstrate that our meta-learning-based model achieves superior noise reduction performance compared to traditional methods.
\end{abstract}



\end{frontmatter}



\section{Introduction}
Active noise control (ANC) is designed to cancel noise through sound superposition \cite{kuo1996active, elliott2000signal, goodwin2010analysis}.
It is proved that ANC has a strong effect on reducing mid-low-frequency noise, which the passive noise control (PNC) approach involving the use of sound barriers and mufflers can not cope with \cite{lee1992compact, tao2021recent, shi2023active}. With the development of computing and processing devices, ANC technology has advanced rapidly. A wide range of applications have been successfully commercialized, such as earphones \cite{zhang2022robust} and vehicles \cite{cheng2024optimal}. A typical ANC system requires a secondary speaker to produce a secondary sound source and an error microphone to measure residual noise. Depending on whether a reference microphone is used, ANC systems can be classified into feedforward control and feedback control \cite{kuo1999active}. If both feedforward and feedback structures are employed, the system is considered hybrid \cite{song2005robust}.

Traditionally, linear adaptive filters are used to handle Gaussian distributed noise passing through the primary acoustic path \cite{ingle2005statisical}. The least mean square (LMS) algorithm and its variants are commonly employed due to their simplicity, robustness, and low computational load \cite{haykin2002adaptive}. The filtered-x LMS (FxLMS) algorithm is proposed \cite{gaur2016review, chen1998stability, manzano2018optimal} to account for the time delay of the secondary path. However, nonlinear distortions are inevitable in real-world ANC implementations. The responses of electronic components, such as loudspeakers, are not strictly linear. In particular, the saturation effect of electronic components mainly distorts the secondary source. Therefore, linear adaptive filtering is unable to achieve perfect ANC modeling.

Researchers have proposed various adaptive filters to address nonlinear distortions in electrical devices and acoustic paths. The leaky-update and weighted-update algorithms are suggested to mitigate the nonlinearity of electrical devices \cite{tobias2002performance, tobias2005leaky}. The Volterra kernel has been introduced to project the reference signal onto a higher-dimensional space to handle nonlinearity in ANC systems \cite{tan1997filtered, tan2001adaptive, zhao2011adaptive}. The tangential hyperbolic function-based FxLMS algorithm models the secondary path with saturation-type nonlinearity \cite{sahib2012nonlinear, ghasemi2016nonlinear}. Beyond nonlinear function-based FxLMS algorithms, neural networks have also been exploited to handle system-wide nonlinearity \cite{zhou2005analysis}. Functional-link structures use nonlinear, fully-connected layers to achieve better performance \cite{behera2014functional, le2018bilinear, zhu2024quantized}. In \cite{snyder1995active}, a multilayer perceptron (MLP) network is adopted for active control of vibrations. Furthermore, various schemes have been developed to increase convergence speed and reduce computational complexity \cite{bouchard1999improved, zhang2006adaptive}. Additionally, robust adaptive filters based on M-estimator and fractional lower-order statistics \cite{kumar2020modified, yang2022fractional, patel2023generalized, feng2023active} have been developed to handle abnormal disturbances and signals.

Deep learning has gained significant traction and is supported by well-developed and accessible databases, such as speech corpus \cite{zue1990speech} and hand gestures \cite{kapitanov2024hagrid}. Deep learning models can be trained and applied in practical scenarios due to sufficient data and computational resources \cite{deng2014deep}. Unlike traditional MLP, convolutional neural networks (CNNs) can compress and extract input features \cite{li2021survey}. Furthermore, recurrent neural networks (RNNs), including long short-term memory (LSTM) models, have proven to be effective in dealing with long data sequences for time series prediction \cite{graves2012long, fu2016using, duan2016travel}. The work \cite{park2019long} integrates MLP, CNN, and RNN in a feedforward ANC system. Moreover, convolutional recurrent network (CRN) structures have been applied to ANC, showing superior performance \cite{zhang2021deep, zhang2023deep}. A deep selective fixed-filter structure is recently developed \cite{luo2023deep}, which models the original ANC problem as an adaptive switching problem. After that, many deep-learning-based methods have been devised \cite{luo2023delayless, oh2024enhancing}.

Apart from the end-to-end deep ANC methods, the deep adaption method has been proposed in \cite{zhang2022deep}. In \cite{casebeer2021auto}, a meta-learning-based adaptive filter is introduced to learn robust updating rules. The work \cite{casebeer2022meta} devises a meta-learning-based adaptive filter employing a fast block updating structure. However, ANC systems involve a physical secondary path and a loudspeaker, implying that fast block updating may violate the time constraint \cite{yang2018frequency}.

In this paper, we devise a meta-learning-based delayless subband ANC architecture to improve the performance of feedforward ANC systems in various noisy and nonlinear environments, including the nonlinearity of the loudspeaker. The modified delay-less subband relaxes the time constraint in the ANC systems. We also discuss a variant of the proposed architecture with the main delay of the secondary path. Extensive simulations under different conditions demonstrate the robustness and efficiency of the meta-learning-based delayless subband ANC architecture and its superiority over traditional ANC algorithms.

The main contributions are summarized as follows:
\begin{itemize}
\item \textbf{Meta-Learning-Based Adaptive Filter for ANC}: We reformulated the ANC problem of applying adaptive filtering as a meta-learning problem to learn an optimized updating rule for adaptive filter from noisy observations. A deep learning model with learnable feature embedding approximates the optimized updating rule.
\item \textbf{Modified Delayless Subband Architecture}: The delayless subband architecture is introduced, which does not demand multi-point updating. We modified the original architecture to incorporate it into the learning-based adaptive filter. The modified architecture decreases the updating frequency to relax the time constraint of updating the adaptive filter weight. Besides, a skip updating strategy is introduced to decrease the update frequency further so that machines with limited computational resources can board the proposed deep learning model.
\item \textbf{Training with Partially Known Secondary Path}: We further modified the proposed model to cope with the situation where only the main delay of the secondary path is known. Training with the main delay of the secondary path,  our model can correctly update the adaptive filter weight.
\item \textbf{Extensive Numerical Simulations}: Large-scaled and multi-conditioned data for training and testing is generated using public databases. The large and complex training set ensures the model learns an updating rule with good generalization and robustness. The testing set verified the proposed model's superior performance and robustness over the traditional adaptive filters. The source code is available at:
\end{itemize}

The remainder of this paper is organized as follows: Section \ref{SEC:PS} describes the signal model of ANC and its linear solution. Section \ref{SEC:PA} develops the meta-learning-based delayless subband ANC architecture and skip updating strategy. Section \ref{SEC:ES} details the experimental setup. Section \ref{SEC:ERC} includes simulation results and discussion. Finally, Section \ref{SEC:CC} concludes the paper.

\section{Problem Statement}
\label{SEC:PS}
\begin{figure}[h]
\centering{
\includegraphics[scale=1]{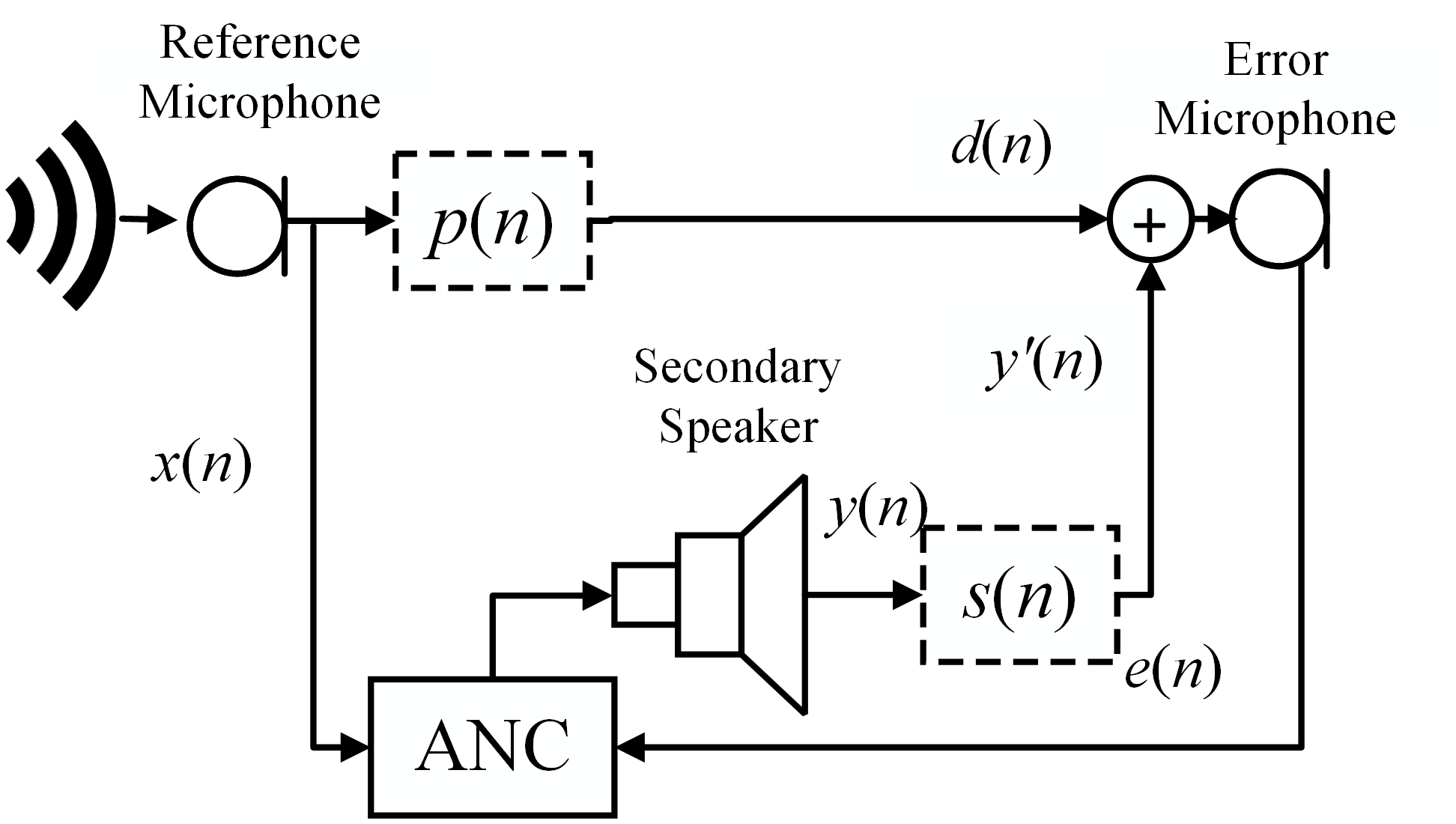}}
\caption{\centering{Diagram of single-channel feedforward ANC system}}
\label{FIG:ANC}
\end{figure}
\subsection{Signal Model}

The block diagram of the ANC system is depicted in Fig. \ref{FIG:ANC}, where $p(n)$ and $s(n)$ denote the primary and secondary acoustic paths, respectively. The primary and secondary paths come from the noise source and the secondary speaker, respectively, to the error microphone. The reference signal $x(n)$ is picked up by
the reference microphone close to the noise source. Without loss of generality, we assume that the reference signal $x(n)$ is equal to the noise source \cite{yang2020stochastic}. Typically, the secondary path $s(n)$ is represented by a finite impulse response (FIR) filter of length $L$, with coefficient vector $\textbf{s}=[s_0, s_1, ..., s_{L - 1}]^T$. The ANC controller generates the secondary sound source $y(n)$ by employing a linear adaptive filter:
\begin{equation}
y(n) = \textbf{w}^T(n)\textbf{x}(n)
\label{EQ:SSS}
\end{equation}
where $T$ is the transpose operator. The adaptive filter weight and the reference vectors are represented as $N$-length vectors $\textbf{w}(n) = [w_0(n), w_1(n), ..., w_{N - 1}(n)]^T$ and $\textbf{x}(n) = [x(n), x(n-1), ..., x(n - N + 1)]^T$, respectively.

This secondary sound source $y(n)$ then passes through the secondary acoustic path, producing the anti-phase sound $y'(n)=s(n)*y(n)$ that cancels the desired signal $d(n)$. The corresponding error signal $e(n)$ can be calculated as:
\begin{equation}
e(n) = d(n) + y'(n) = d(n) + s(n) * y(n)
\label{EQ:ERR}
\end{equation}
where $*$ denotes the convolution operator.

Considering the nonlinear saturation effect of the secondary loudspeaker, a nonlinear function $f_{\mathrm{NSE}}$ can be used \cite{tobias2006lms}:
 \begin{equation}
f_{\mathrm{NSE}}[y] = \int_0^y  \exp\left[\frac{z^2}{2\eta^2}\right]dz
\label{EQ:SEF}
\end{equation}
where $\eta$ controls the nonlinearity. The smaller the value of $\eta$ is, the higher nonlinearity the loudspeaker has. If $\eta^2\rightarrow\infty$, (\ref{EQ:SEF}) reduces to $f_{\mathrm{NSE}}[y]\rightarrow y$.

Using (\ref{EQ:SEF}), (\ref{EQ:ERR}) can be rewritten as:
 \begin{equation}
e(n) = d(n) + s(n) * f_{\mathrm{NSE}}[y(n)]
\label{EQ:ERR_2}
\end{equation}

\subsection{Linear Solution} 

\begin{figure}[h]
\centering{
\includegraphics[scale=1]{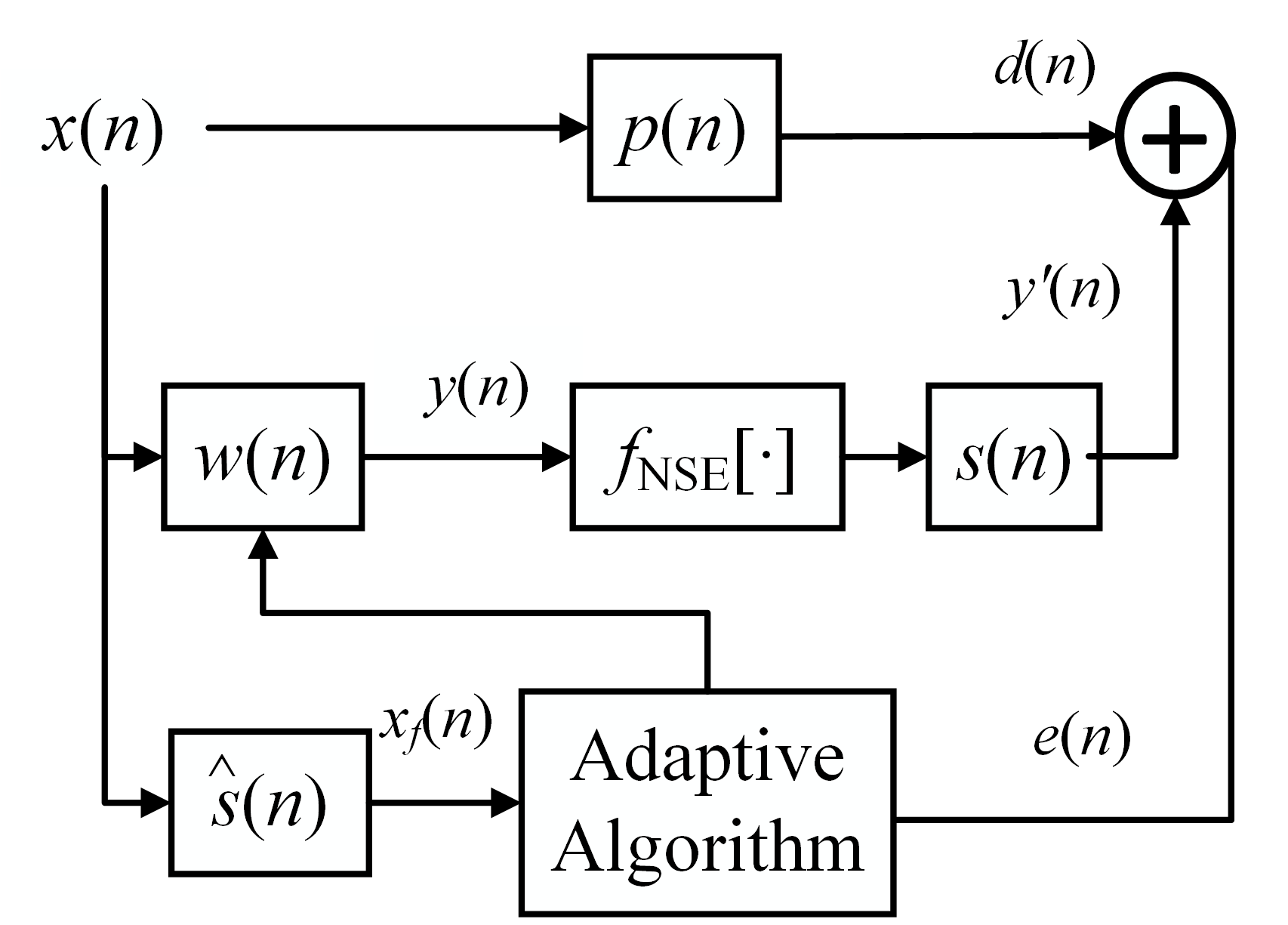}}
\caption{\centering{Diagram of ANC system with adaptive algorithm}}
\label{FIG:LANC}
\end{figure}

Fig. \ref{FIG:LANC} shows the traditional linear adaptive filter applying in the single-channel ANC system. An estimated secondary path $\hat{s}(n)$ is used for aligning signals in the time domain. The secondary path is assumed to be precisely identified. In this case, the filtered reference signal can be obtained as:
\begin{equation}
x_f(n) = \hat{s}(n) * x(n)
\label{EQ:XF}
\end{equation}

By ignoring the nonlinearity of the speaker, the optimal linear solution can be obtained by minimizing the mean squared error (MSE) function $E\{e^2(n)\}$. In practice, we minimize the instantaneous squared error signal $e^2(n)$ instead of $E\{e^2(n)\}$. Using (\ref{EQ:ERR_2}) and (\ref{EQ:SSS}), the resultant algorithm is:
\begin{equation}
\textbf{w}(n+1)=\textbf{w}(n) - \mu \textbf{x}_f(n)e(n)
\label{EQ:LMS}
\end{equation}
where $\mu > 0$ is the step size. 

By analyzing the traditional adaptive updating rule in (\ref{EQ:LMS}), the convergence depends on the upper bound of the step size $\mu$, which is negatively correlated with $||\textbf{x}(n)||_2^2$ \cite{sun2015convergence} and the delay of the secondary path \cite{ardekani2010theoretical}. Suppose an abnormal component in the reference signal $\textbf{x}(n)$ exists. In that case, the upper bound will be extremely small to maintain stability, rendering the adaptive filter ineffective for the corresponding system. Additionally, when the disturbance is nonstationary, the error signal $e(n)$ will destabilize the adaptive filter. Here, we reconsider the nonlinear saturation effect of the loudspeaker. The loudspeaker's nonlinear saturation effect limits the secondary source's amplitude. As $\eta$ decreases, the maximum amplitude of the secondary source becomes smaller and smaller, and the residual noise becomes larger and larger, which will prevent the traditional adaptive filter from achieving perfect modeling.

\section{Proposed Algorithm}
\label{SEC:PA}

\subsection{Meta Learning}
Meta learning can be defined as improving a learning algorithm through learning episodes \cite{hospedales2021meta}. It involves a base learner and a meta learner. The meta learner is designed to acquire meta knowledge across different datasets, enabling the base learner to adapt to new tasks \cite{finn2019online} quickly. In ANC systems, the base learner functions as the current adaptive filter weight, while the meta learner represents the adaptive updating rule for the filter \cite{casebeer2022meta}. The meta learner predicts the gradient $\tilde{\textbf{g}}(n)$ to update the adaptive filter weight. If the meta learner employs the FxLMS algorithm, the predicted gradient is:
\begin{equation}
\tilde{\textbf{g}}(n) =  \textbf{x}_f(n)e(n)
\label{EQ:UP_LMS}
\end{equation}

Here, we consider a neural network (NN) model with a temporal hidden state $\textbf{h}$ and modify (\ref{EQ:LMS}) as:
\begin{equation}
\textbf{w}(n + 1) = \textbf{w}(n) - \mu \tilde{\textbf{g}}_\phi[\cdot, \textbf{h}]
\label{EQ:UP_M}
\end{equation}
where $\tilde{\textbf{g}}_\phi[\cdot, \textbf{h}]$ is the instantaneous output of $\tilde{\textbf{g}}_\phi(n)$ whose parameters are $\phi$.

\begin{figure}
\centering{
\includegraphics[scale=1]{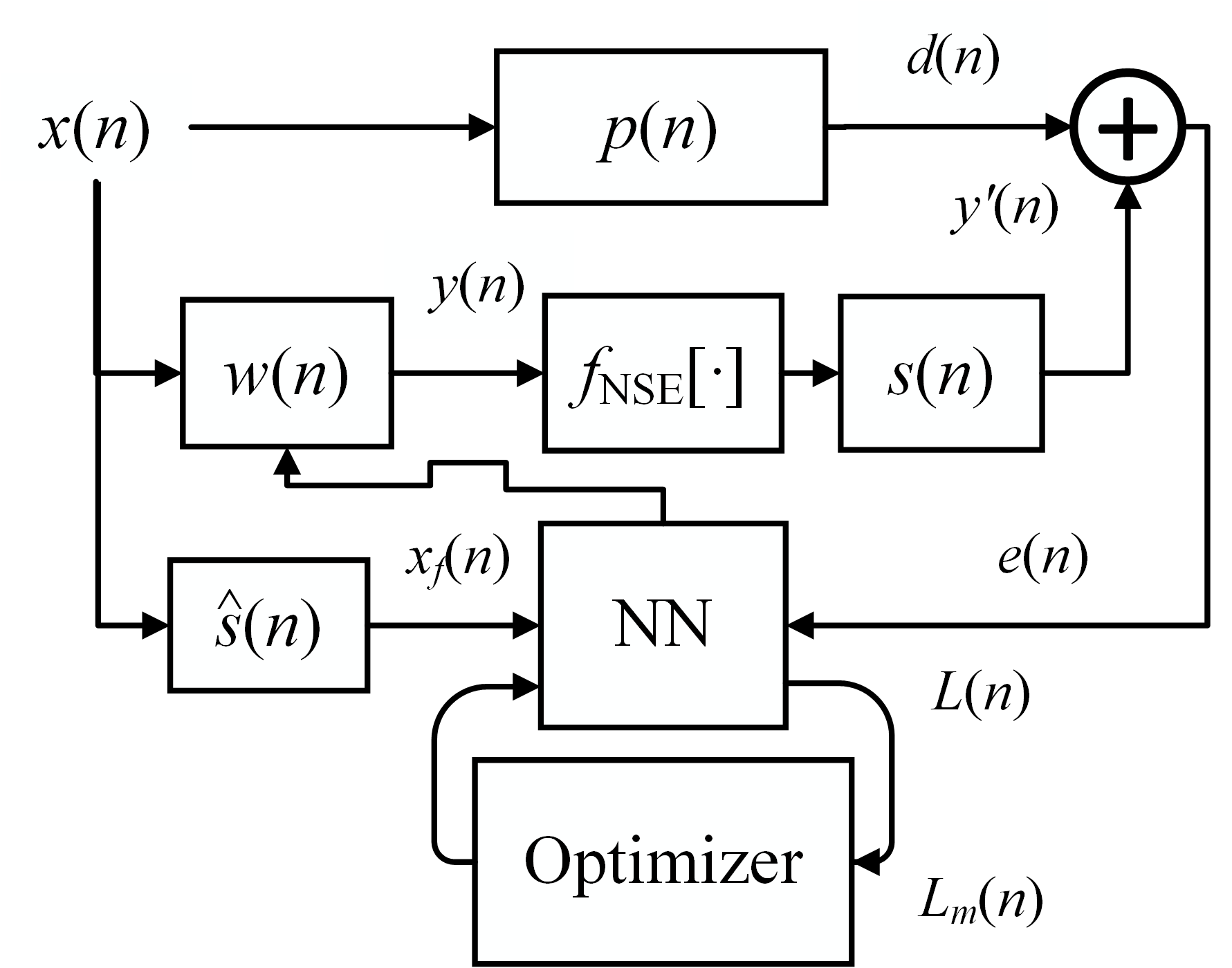}}
\caption{\centering{Diagram of ANC system with NN model using meta learning}}
\label{FIG:META}
\end{figure}

Contrast (\ref{EQ:UP_M}) with (\ref{EQ:LMS}), it can be seen that $\tilde{\textbf{g}}_\phi[\cdot, \textbf{h}]$ is the predicted gradient. In this case, (\ref{EQ:UP_M}) is the optimized updating rule in adaptive filters. Then, an optimal adaptive updating rule is searched by training the NN model across the dataset $\mathcal{D}$ \cite{casebeer2022meta}:
 \begin{equation}
\hat{\phi} = \mathop{\arg}\mathop{\min}_\phi E_\mathcal{D} \{\mathcal L_M[\mathcal L(n),\tilde{\textbf{g}}_\phi(n)] \}
\label{EQ:UP_MM}
\end{equation}
where $\mathcal L_M$ is the meta loss of the meta learner, $\mathcal L(n)$ is the loss of the base learner at time $n$ and $\tilde{\textbf{g}}_\phi(n)$ is the instantaneous predicted gradient of the NN model . A simple training scheme is illustrated in Fig. \ref{FIG:META}. The training data is typically extensive, so the NN model is optimized batch by batch. In addition, the loss of the base learner $\mathcal L(n)$ is $\ell_2$ norm loss, and the meta loss $\mathcal L_M = \sum{\mathcal L(n)}$. It is worth mentioning that $\mathcal L(n)$ is computed from noisy observations. Whenever receiving a error signal $e(n)$, the NN model will update adaptive filter weight and compute the instantaneous loss $\mathcal L(n)$. After one batch of training, a gradient decent-based optimizer will be used to optimize the parameters of the NN model with $E_\mathcal{D} \{\mathcal L(n)\}=E_\mathcal{D} \{\sum{\mathcal L(n)}\}$. After several training epochs, the meta loss converges. Then, the NN model can predict a reasonable gradient to update the adaptive filter weight. 

\subsection{Modified Delayless Subband Architecture}
Recalling the deep feedforward ANC approach, \cite{park2019long} employs various deep learning modules, such as CNNs, to predict the desired signal $d(n)$ in the time domain. The approach of deep learning modules is similar to MLP \cite{tokhi1997active}. However, the performance of these simple models is questionable when the primary path changes. In another study, \cite{zhang2021deep} uses spectra in both the input and output, predicting the desired signal $d(n)$ with the overlap-add method, which requires fitting multiple points. It is worth mentioning that using a multiple-point frequency domain algorithm \cite{zhang2021deep} will introduce an extra time delay in the ANC process, which may violate the time constraint \cite{yang2018frequency}.

To mitigate the time delay problem and make the deep learning-based adaptive filter practical, we use a modified delayless subband architecture as the backbone of our learning-based model. The original delayless subband adaptive filter was developed in \cite{morgan1995delayless}. Its main idea is to stack weight vectors in subbands to produce the fullband filter vector, thereby alleviating the time alignment problem.

\begin{figure}
\centering{
\includegraphics[scale=0.9]{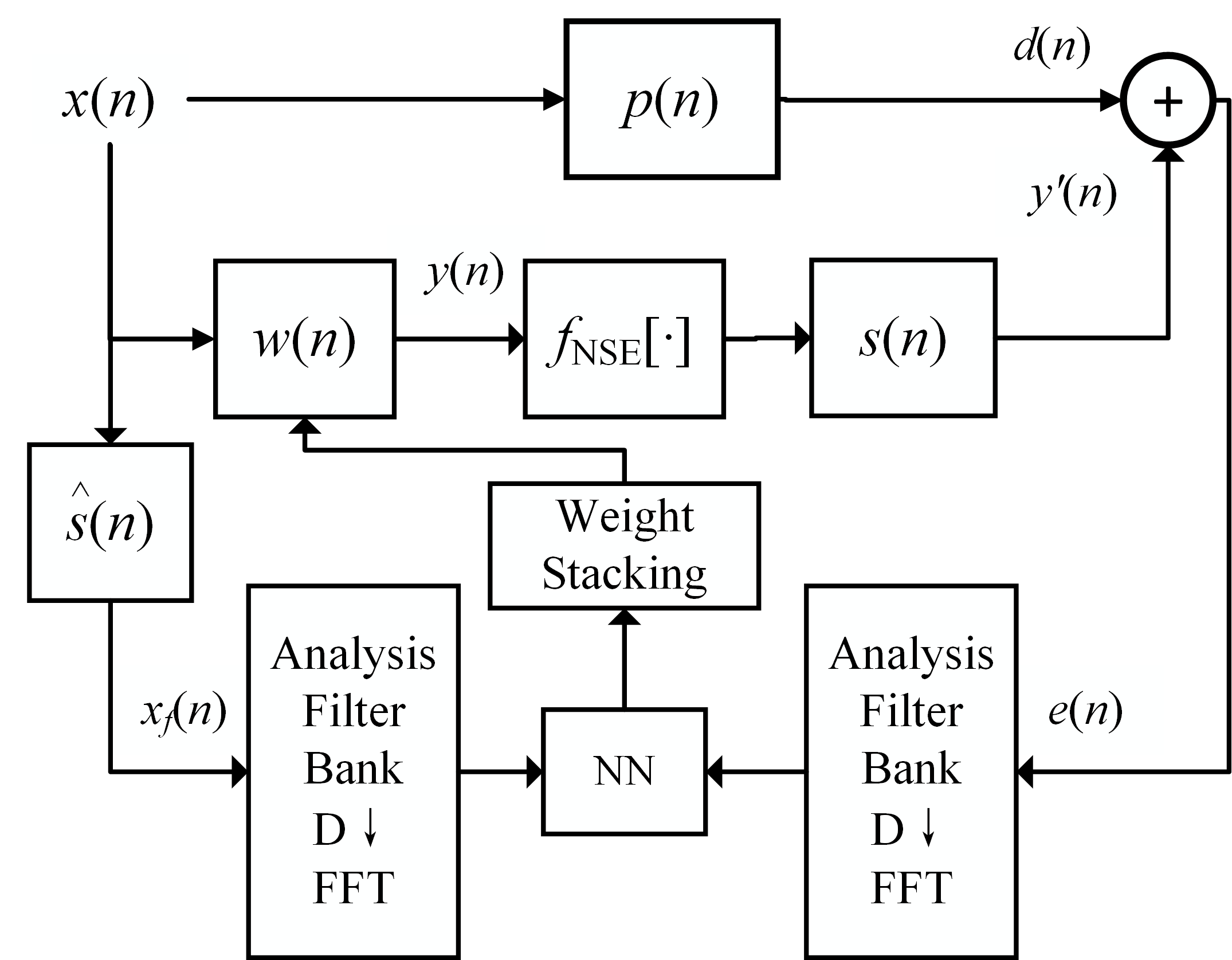}}
\caption{\centering{Diagram of modified delayless subband architecture with NN model}}
\label{FIG:DSANC}
\end{figure}

In the original delayless subband architecture, the filtered reference signal $x_f(n)$ and the error signal $e(n)$ are decomposed into several subband signals by an analysis filter bank with a downsampling factor $D$. Here, we use the polyphase technique to generate the analysis filter bank, whose coefficients are computed as follows:
\begin{equation}
\begin{cases}
\textbf{a}_k = [a^0_k, a^1_k, ..., a^{Q-1}_k]^T\\
a^q_k = c_q \exp[\mathrm{j} 2\pi qk/K]
\end{cases}
\label{EQ:AFB}
\end{equation}
where $c_q$ represents the $q$th coefficient of the prototype filter, $K$ is the subband number, $\mathrm{j}$ is the imaginary unit, $D = K/2$ and $Q=N/D$ is the subband filter length. 

Then, the resultant filtered subband signals are obtained as:
\begin{equation}
\begin{cases}
x_{fk}(n) = \textbf{a}^T_k \textbf{x}_{fs}(n) \\
e_{k}(n) = \textbf{a}^T_k \textbf{e}_{s}(n)
\end{cases}
\label{EQ:SXE}
\end{equation}
where $\textbf{x}_{fs}(n) = [x_f(n), x_f(n - D), ..., x_f(n - QD + D)]^T$ and $\textbf{e}_s(n) = [e(n), e(n - D), ..., e(n - QD + D)]^T$.

It is worth mentioning that a smaller adaptive filter update frequency is feasible due to the presence of the analysis filter bank. For example, if the sampling frequency used is 16 kHz and the downsampling factor $D$ is $16$, the system will be able to operate at a frequency of $1$ kHz. 

To form the fullband filter weight using the subband filter weights, the frequency stacking method is required. Typically, the fast Fourier transform (FFT)-1 method \cite{li2024enhanced} is adopted:
\begin{equation}
\begin{cases}
\textbf{w}^{(l)}_{f}(n) = 
\begin{cases}
\textbf{w}^{(l)_{2N/K}}_{f\lfloor lK/N \rfloor}(n), l \in [0, N/2) \\
0,  l = N/2 \\
\textbf{w}^{(N-l)}_{f}(n), l \in (N/2, N) \\
\end{cases} \\
\textbf{w}_{f}(n)=[\textbf{w}_f^{(0)}(n), \textbf{w}_f^{(1)}(n), ..., \textbf{w}_f^{(L-1)}(n)]^T
\end{cases}
\label{EQ:TWS}
\end{equation}
where $\textbf{w}^{(l)}_{f}(n)$ is the $l$th coefficient of the fullband filter in the frequency domain, $\textbf{w}^{(l)}_{fk}(n)$ means the $l$th coefficient of the $k$th subband filter in the frequency domain, $\lfloor \cdot \rfloor$ is the floor operator and $(\cdot)_{2N/K}$ is the modulo-$2N/K$ operator.

Our modified architecture is depicted in Fig. \ref{FIG:DSANC}. The core updating rule is represented by an NN model. Different from the traditional delayless subband structure, FFT is directly used after the analysis filter bank to generate more stable features of the filtered subband reference signal $x_{fk}(n)$ and the subband error signal $e_k(n)$. The resultant filtered subband signals are written as:
\begin{equation}
\begin{cases}
\textbf{x}_{ffk}(n) = \mathrm{FFT}[x_{fk}(n), x_{fk}(n-1),
 ..., x_{fk}(n-Q+1)] \\
\textbf{e}_{fk}(n) = \mathrm{FFT}[0, 0, ..., 0, \textbf{e}_{k}(n)]
\end{cases}
\label{EQ:SBXE}
\end{equation}
where there are $Q-1$ zeros padded for computing $\textbf{e}_{fk}(n)$ and time alignment.

Noting the symmetry of the weight vector in the frequency domain, we only need to update half of the subband weights in traditional subband adaptive filter.  In our architecture, we directly update the subband adaptive filter weights $\textbf{w}_s(n)$ in the frequency domain:
\begin{equation}
\textbf{w}_s(n + 1) = \textbf{w}_s(n) - \mu \tilde{\textbf{g}}_\phi[\cdot, \textbf{h}]
\label{EQ:UP}
\end{equation}

Distinguishing from the traditional weight stacking method in (\ref{EQ:TWS}), we directly stack the weights $\textbf{w}_s(n)$ to obtain $\textbf{w}(n)$. The stacking rule is given as:
\begin{equation}
\textbf{w}^{(l)}_f(n) = 
\begin{cases}
\textbf{w}^{(l)}_{s}(n), l \in [0, N/2) \\ 
0, l = N/2 \\
\textbf{w}^{(N-l)}_{s}(n), l \in (N/2, N) \\
\end{cases} \\
\label{EQ:FFF}
\end{equation}

Then, we directly use IFFT to reconstruct the fullband adaptive filter weight:
\begin{equation}
\textbf{w}(n) = \mathrm{IFFT}[\textbf{w}_f(n)]
\label{EQ:FFFF}
\end{equation}

\subsection{Neural Network with Complex Self-Attention}
\label{SUB:CA}

In this subsection, we devise the core module to update the subband filter weights.

\begin{figure*}
\centering{
\includegraphics[scale=0.8]{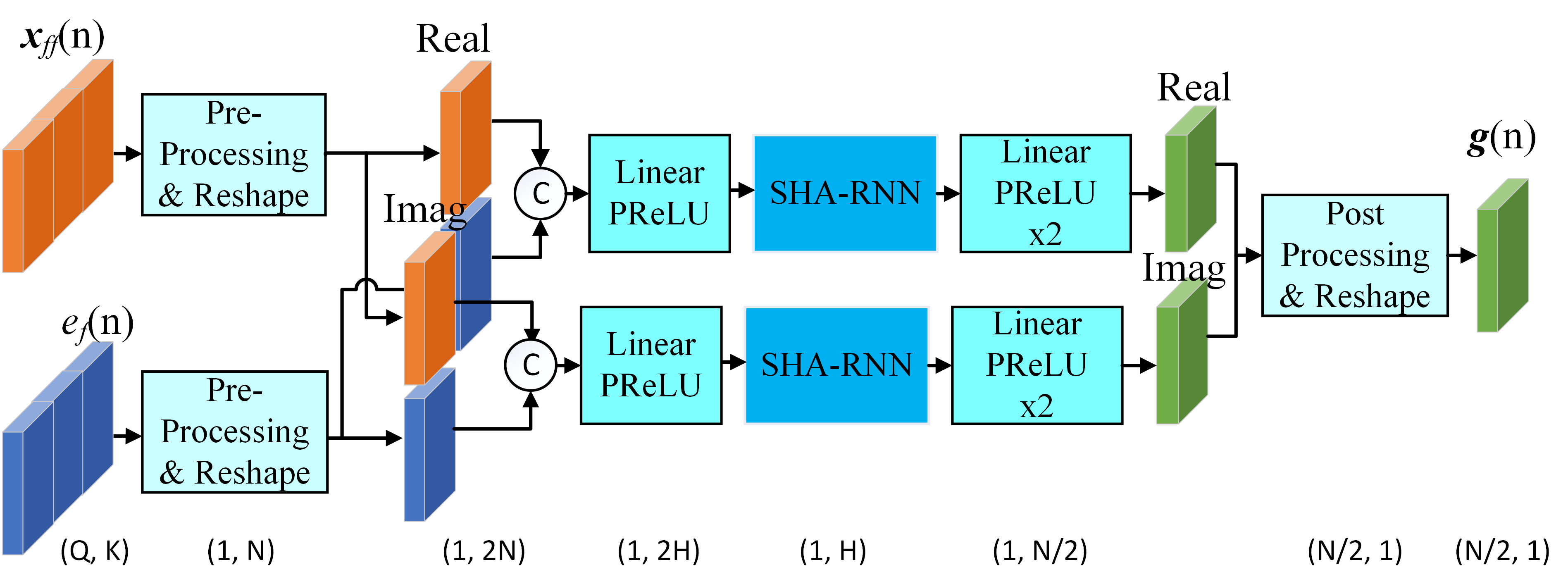}}
\caption{\centering{Proposed neural network architecture with complex self-attention, (x, y) means the data's dimensions after operations}}
\label{FIG:NN}
\end{figure*}

In \cite{casebeer2022meta}, a small deep learning model is used as the core updating rule. This module consists of three layers of fully-connected feedforward networks and two layers of gated recurrent unit (GRU) networks. When updating the weight vector, the model in \cite{casebeer2022meta} updates the vector in each frequency bin, which makes the method relatively slow. In ANC system, we must address the issues of time alignment and time delay. Therefore, we opt to update the first half of the fullband adaptive filter weights $\textbf{w}_s(n)$ in a single output. As shown in Fig. \ref{FIG:NN}, the input to the NN consists of half of the filtered subband reference signals and the error signals. The number of intermediate units is fixed at $H$. Before feeding the inputs into the NN model, the input complex signals $\textbf{x}_{ffk}(n)$ and $\textbf{e}_{fk}(n)$ are compressed by a nonlinear function as:
\begin{equation}
\mathrm{ln}[1 + |\beta|]e^{j\angle\beta}
\label{EQ:LN}
\end{equation}
where $\beta$ is the input complex value. In practice, this nonlinear function is not as useful as it is in work \cite{andrychowicz2016learning} and does not improve performance. Here, we still apply (\ref{EQ:LN}) to our model because it compresses the dynamic range of the input value and accelerates the training phase. 

\begin{figure*}
\centering
\subfigbottomskip=1pt 
\subfigcapskip=1pt
\subfigure[Details of SHA-RNN module]{
\includegraphics[scale=1]{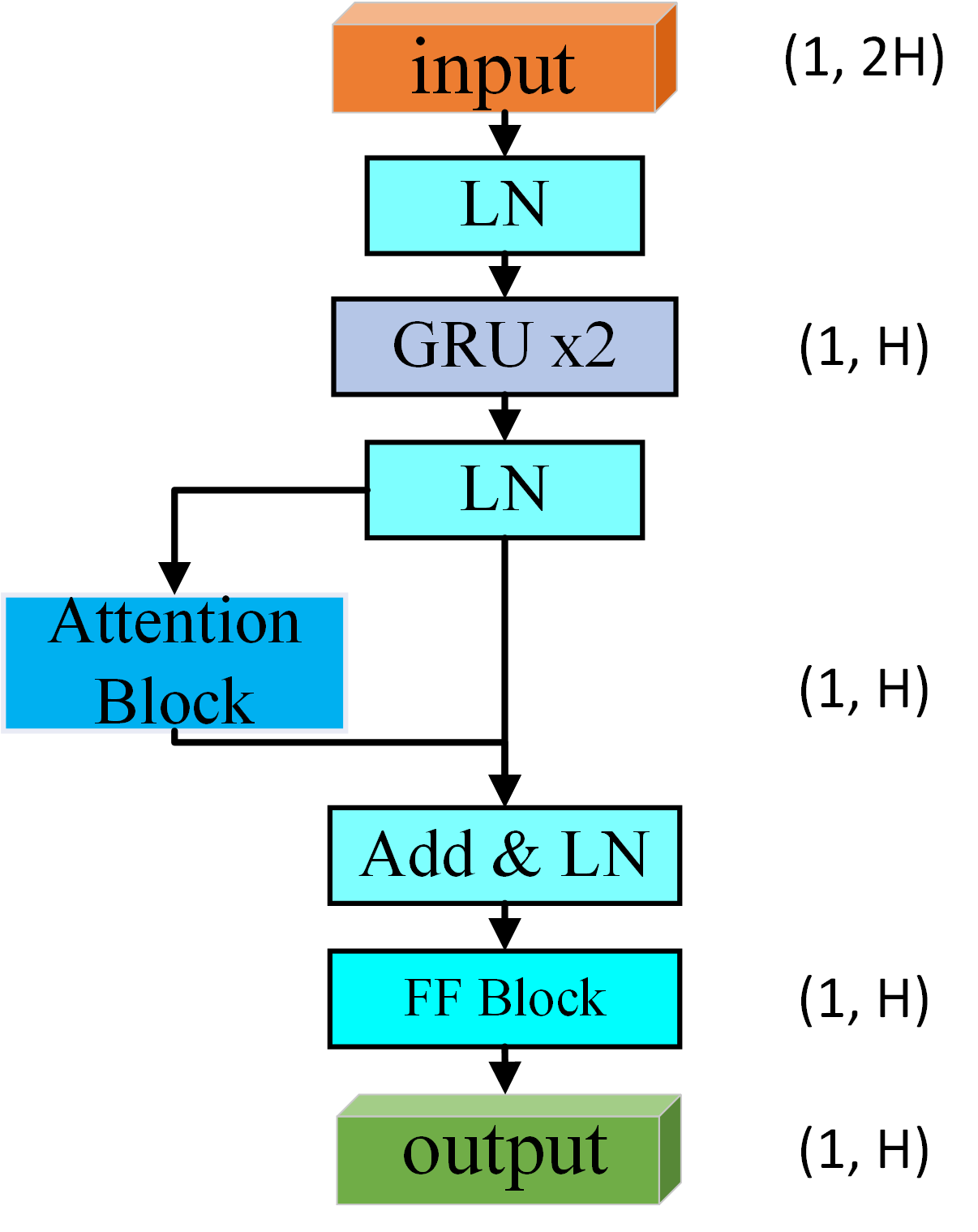}}
\hspace{0cm}
\subfigure[Details of proposed attention block]{
\includegraphics[scale=1]{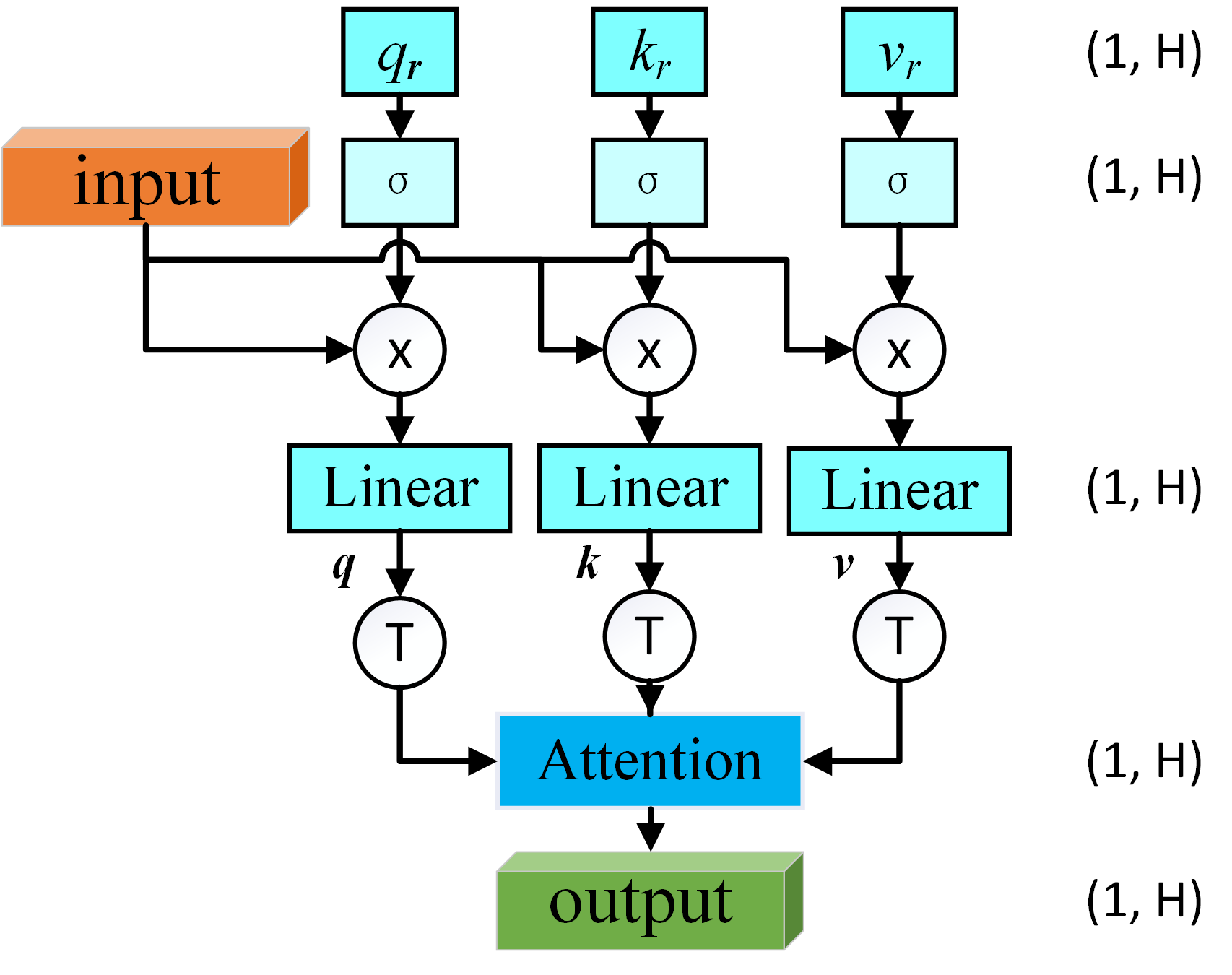}}
\caption{\centering{Proposed SHA-RNN module with learnable feature embedding , (x, y) means the data's dimensions after operations}}
\label{FIG:BLOCK}
\end{figure*}

When inputting signals into our model, we use a fully-connected feedforward network layer as the basis generator. The input consists of the concatenated real and imaginary parts of the filtered subband reference signal $\textbf{x}_{ffk}(n)$ and the subband error signal $\textbf{e}_{fk}(n)$ in the frequency domain. Then, a single-headed attention RNN (SHA-RNN) block is utilized to process the features. The SHA-RNN is proposed in \cite{merity2019single}, whose structure is depicted in Fig. \ref{FIG:BLOCK}(a), where the LN block stands for layer normalization. The feedforward block \cite{vaswani2017attention} relays the over-fitting problem. Because we have yet to modify the original structure of the feedforward block, its details will not be discussed in this paper. 

The original SHA-RNN does not specify the calculation method of the attention mechanism. Hence, we can design our attention mechanism. Considering that although GRU has processed the filtered subband signals before the attention block, the position of the filtered subband signals comprises the temporal information. Inspired by \cite{vaswani2017attention}, which uses an input positioning method, we propose a learnable feature embedding mechanism and combine it into the attention block as shown in Fig. \ref{FIG:BLOCK}(b). Here, the input dimension is $H$, so the dimensions of the learning parameters $Q_r$, $K_r$, and $V_r$ are all $H$. The expressions of the inquiry vector $\textbf{q}$, key vector $\textbf{k}$ and the value vector $\textbf{v}$ can be written as: 
\begin{equation}
\begin{cases}
\textbf{q} = \mathrm{Linear}[\cdot \odot \sigma[\textbf{q}_r]] \\
\textbf{k} = \mathrm{Linear}[\cdot \odot \sigma[\textbf{k}_r]] \\
\textbf{v} = \mathrm{Linear}[\cdot \odot \sigma[\textbf{v}_r]])
\end{cases}
\end{equation}
where $\odot$ is the element-wise multiplication operator, $\sigma[\cdot]$ is the sigmoid function and $\mathrm{Linear}[\cdot]$ represents linear function.

Once we have the required $\textbf{q}$, $\textbf{k}$ and $\textbf{v}$, we perform the calculation of attention:
\begin{equation}
\mathrm{Softmax} [\textbf{q}^T \textbf{k}]\textbf{v}^T
\label{EQ:ATT}
\end{equation}
where $\mathrm{Softmax}[\cdot]$ is the softmax function. It is worth mentioning that (\ref{EQ:ATT}) is different from the original one. In the original attention calculation, all the $\textbf{q}$, $\textbf{k}$ and $\textbf{v}$ are matrices whose columns are temporal axis and the rows are feature axis. Here, we swapped the dimension of features with the dimension of time to let the $\textbf{q}_r$, $\textbf{k}_r$, $\textbf{v}_r$ learning the feature weights.

After SHA-RNN processing, two layers of fully-connected networks are used as the decoder. In addition, parametric rectified linear unit (PReLU) \cite{he2015delving} is adopted as the activation function after each linear layer, excluding the linear layer in the attention block. Noticing that the output of the NN model is actually a gradient in the frequency domain, we postprecess the output to constrain its amplitude:
\begin{equation}
\begin{split}
\tilde{\textbf{g}}(n) = & \left(\ln\left[ \frac{\mathrm{max}[\mathrm{min}[|\textbf{g}(n)|, \exp[-10]], \exp[10]]}{10}\right] + 1\right)\exp[j\angle\textbf{g}(n)]
\end{split}
\label{EQ:TH}
\end{equation}
where $\textbf{g}(n)$ is the original predicted gradient of the NN model. (\ref{EQ:TH}) actually limits the amplitude of the predicted gradient from $0$ to $2$, which is actually the ideal step size interval of the normalized FxLMS (NFxLMS).

\begin{table}
\caption{Training algorithm}
\label{TB:TA}
\begin{tabular}{l}
\toprule
$\textbf{Initialization}$: $\mathrm{NN}[\cdot]$, $\textbf{w}(0)$, $F$, $\textbf{a}$, $D$, $\mu$\\
\midrule
$\textbf{Function}$: Inner loop$(\textbf{w}(n), \textbf{x}, \textbf{d},\mathrm{NN}[\cdot], F, N, \textbf{h}, D, \textbf{a}), \mu$ \\
$\quad$ $\mathcal L_M = 0$ \\
$\quad\textbf{For} \;\; i = 0, 1, 2, ..., FD-1$ \\
$\quad\quad$ $y(n) =\textbf{w}^T(n+i)\textbf{x}(n+i)$ \\
$\quad\quad$ $y'(n) =f_{\mathrm{NSE}}^T[\textbf{y}(n+i)]\textbf{s}$ \\
$\quad\quad$ $x_f(n) =\textbf{x}^T(n+i)\hat{\textbf{s}}$ \\
$\quad\quad$ Compute $\textbf{x}_{ffk}(n+i)$ and $\textbf{e}_{fk}(n+i)$ using (\ref{EQ:AFB}), (\ref{EQ:SXE}) and (\ref{EQ:SBXE})\\
$\quad\quad$ $\textbf{If}\;\;n \% D == 0$\\
$\quad\quad\quad$ $\tilde{\textbf{g}}(n+i), \textbf{h} =\mathrm{NN}[\mathrm{Re}[\textbf{x}_{ff0}^T(n+i)], \mathrm{Re}[\textbf{e}_{f0}^T(n+i)], ...,$ \\
$\quad\quad\quad\quad$ $\mathrm{Im}[\textbf{x}_{ff(Q/2-1)}^T(n+i)], \mathrm{Im}[\textbf{e}_{f(Q/2-1)}^T(n+i)], \textbf{h}]$ \\
$\quad\quad\quad$ $\textbf{w}_s(n+1+i) = \textbf{w}_s(n+i) - \mu \tilde{\textbf{g}}(n+i)$\\
$\quad\quad\quad$ Compute $\textbf{w}(n+i)$ using (\ref{EQ:FFF})\\
$\quad\quad$ Compute $\mathcal L(n+i)$ using (\ref{EQ:LOSS}) \\
$\quad\quad\mathcal L_M = \mathcal L_M + \mathcal{L}(n+i)$ \\
$\quad\textbf{End for}$\\
$\quad\textbf{Return}\;\mathcal L_M, \textbf{w}(n+i)$ \\
$\textbf{End function}$ \\
$\textbf{Function}$: Outer loop($\textbf{w}(0)$, $\mathrm{NN}[\cdot]$, $F$, $D$, $\textbf{a}$, $\mu$) \\
$\quad$Extract $\textbf{x}$, $\textbf{d}$, $\textbf{s}$ from $D$\\
$\quad \textbf{While not converged do}$\\
$\quad\quad \mathcal L_M, \textbf{w}(n+FD) =\;$Inner loop$(\textbf{w}(n), \textbf{x}, \textbf{d},\mathrm{NN}[\cdot], F, N, \textbf{h}, \textbf{a}, \mu)$\\
$\quad\quad\textbf{h}=$ Meta optimization $ (\mathcal L_M,  \textbf{h})$\\
$\quad\quad n=n+FD$\\
$\quad \textbf{End\;while}$\\
$\quad\textbf{Return}\;\mathrm{NN}[\cdot]$\\
$\textbf{End function}$ \\
\bottomrule
$\%$ denotes the modulo operator.
\end{tabular}
\end{table}

In summary, we rewrite the relationship between the output $\tilde{\textbf{g}}(n)$ in (\ref{EQ:UP}) and the input signals as:
\begin{equation}
\begin{split}
\tilde{\textbf{g}}(n) = \mathrm{NN}[& \mathrm{Re}[\textbf{x}_{ff0}^T(n)], \mathrm{Re}[\textbf{e}_{f0}^T(n)], \mathrm{Re}[\textbf{x}_{ff1}^T(n)], ..., \\
& \mathrm{Re}[\textbf{x}_{ff(Q/2-1)}^T(n)], \mathrm{Re}[\textbf{e}_{f(Q/2-1)}^T(n)],  \\ 
&\mathrm{Im}[\textbf{x}_{ff0}^T(n)], \mathrm{Im}[\textbf{e}_{f0}^T(n)], \mathrm{Im}[\textbf{x}_{ff1}^T(n)], ..., \\
& \mathrm{Im}[\textbf{x}_{ff(Q/2-1)}^T(n)], \mathrm{Im}[\textbf{e}_{f(Q/2-1)}^T(n)], \textbf{h}]
\end{split}
\label{EQ:NN_UP}
\end{equation}
where $\mathrm{NN}[\cdot, \textbf{h}]$ is the whole function of the proposed NN model.

\subsection{Training Strategies and Loss Function}
In ANC systems, estimating the secondary acoustic path is crucial. For example, the FxLMS algorithm is derived by the gradient descent method, which heavily depends on the estimated secondary path to keep convergent \cite{wu2008improved}. If the estimated secondary path differs from the actual secondary path, FxLMS will slow its convergence process and even become divergent. Since our model is learning-based, we try to train our model without estimating the secondary path. However, our model can only handle the secondary path-free problem when the number of its parameters is vast, which is unsuitable for running in real time. Inspired by work \cite{gao2016simplified}, we attempt to train our model with the main delay of the secondary path. Thus, we choose to train and test our model with the estimated secondary path and with the main delay of the secondary path. Using the main delay of the secondary path, (\ref{EQ:NN_UP}) becomes:
\begin{equation}
\begin{split}
\tilde{\textbf{g}}(n) = \mathrm{NN}[& \mathrm{Re}[\textbf{x}_{\Delta f 0}^T(n)], \mathrm{Re}[\textbf{e}_{f0}^T(n)], \mathrm{Re}[\textbf{x}_{\Delta f 1}^T(n)], ..., \\
& \mathrm{Re}[\textbf{x}_{\Delta f(Q/2-1)}^T(n)], \mathrm{Re}[\textbf{e}_{f (Q/2-1)}^T(n)],  \\
 &\mathrm{Im}[\textbf{x}_{\Delta f0}^T(n)], \mathrm{Im}[\textbf{e}_{f 0}^T(n)], \mathrm{Im}[\textbf{x}_{\Delta f1}^T(n)],  ..., \\
& \mathrm{Im}[\textbf{x}_{\Delta f(Q/2-1)}^T(n)], \mathrm{Im}[\textbf{e}_{f (Q/2-1)}^T(n)], \textbf{h}]
\end{split}
\label{EQ:NN_UP_NS}
\end{equation}
where the subscript $\Delta$ means that the reference signal is filtered by the main delay of the secondary path.

To specify the loss function and the meta loss function in the proposed model, we use the squared error in the frequency domain as the loss function:
\begin{equation}
\mathcal L(n)=||\mathrm{FFT}[0, 0,... e(n)]||^2
\label{EQ:LOSS}
\end{equation}
and the meta loss function is defined as the accumulated MSE:
\begin{equation}
\mathcal L_M=\frac{\sum_{n=0}^{FD-1}\mathcal L(n)}{FD}
\label{EQ:M_LOSS}
\end{equation}
where $F$ is the number of meta frames. Once we obtain the meta loss, gradient decent approach is used to optimize the parameters of the NN model.

Here, we select the best optimizer for our model from $3$ gradient decent-based optimization methods, Adagrad \cite{duchi2011adaptive}, RMSprop \cite{tieleman2012lecture} and ADAM \cite{kingma2014adam}. The ADAM method was found to perform the best in our task, so it is used for further training and testing. The learning rate is set as $0.0001$. The reason is that the meta-optimization will be divergent when the learning rate is too large, while a learning rate that is too small slows down the training. Furthermore, we introduce a learning rate decay scheme to the optimizer. Whenever the new validation loss exceeds the lowest, the learning rate is reduced by $50\%$. 

In summary, we show a simplified form of the training algorithm in Table \ref{TB:TA} with a batch size of $1$. In the following experiments,
we use meta delayless subband adaptive filter (MDSAF) and MDSAF-MD to represent our ANC solutions trained with the whole secondary path and the main delay of the secondary path, respectively.

\subsection{Skip Updating}

Recalling the feedforward ANC system in Fig. \ref{FIG:ANC}, the distance between the noise source and the error microphone should be greater than between the secondary speaker and the error microphone to satisfy the causality constraint. That is, the sound propagation time from the reference microphone to the error microphone $T_p$ should be always larger than the sound propagation time from the secondary speaker to the error microphone $T_s$. As a result, the processing delay of the ANC system $T_{\mathrm{ANC}}$ should be smaller than the difference between $T_p$ and $T_s$:
\begin{equation}
T_{\mathrm{ANC}}  < T_p - T_s
\label{EQ:TC}
\end{equation}

Here, $T_{\mathrm{ANC}}$ includes the time involved in the ADC, convolution of the adaptive filter, DAC, amplifier and secondary speaker. Furthermore, the fullband FxLMS algorithm requires instantaneous gradient computation on a sample-by-sample basis. Hence, the delay in the filtered reference signal is unavoidable. In updating the adaptive filter weight, its processing delay, denoted by $T_{u}$, should be smaller than the sampling interval $\frac{1}{f_s}$:
  \begin{equation}
T_u  < \frac{1}{f_s} < T_p - T_s
\label{EQ:TCI}
\end{equation}

In our approach, computational latency is still unavoidable, but the time constraint of updating can be relaxed by the delayless subband architecture. The resultant time limit for updating is given as:
\begin{equation}
T_{u}  < \frac{D}{f_s}
\label{EQ:TCID}
\end{equation}

Because $\frac{1}{f_s} < \frac{D}{f_s}$, using delayless subband architecture   increase the time limitation to update the adaptive filter weight. Here, we introduce a skip updating factor $B\in \mathbb{Z}^{+}$ to further relax the time constraint: 
\begin{equation}
T_{u}  < \frac{(B+1)D}{f_s}
\label{EQ:TCIP}
\end{equation}

Noting that $\frac{D}{f_s} \leq \frac{(B+1)D}{f_s}$, adjusting the value of $B$ controls the time limitation to update the adaptive filter weight. When $T_{u} < \frac{D}{f_s}$, we can set $B$ as $0$ so that (\ref{EQ:TCIP}) will reduce to (\ref{EQ:TCID}), implying that there is no skip updating. If (\ref{EQ:TCID}) is not satisfied, we can adjust $B$ until (\ref{EQ:TCIP}) is satisfied so that updating the adaptive filter weight will not violate the time constraint. For example, when $T_{u}$ is $1.5$ ms while $\frac{D}{f_s}$ is $1$ ms, $B$ will be set as $1$, meaning that one update will be skipped. 

\subsection{Computational Complexity}

In this subsection, we will summarize the computational complexity of our proposed NN model and then show that it can run in real time.

In our model, the adaptive filter length is $N$, the subband number is $K$, 
 the downsampling factor is $D=K/2$ and the subband filter length is $Q=N/D$.
Then, the length of the input vector of the NN model is $M=2DQ=2N$. Noting that $M=2DQ=2N$, the number of the subband number does not affect the complexity of the NN model. In addition, the number of hidden state units is $H$ and the length of the output vector is $Z=DQ/2=N/2$. In this case, the complexities of the first fully-connected layer, the last two fully-connected layers and the feedforward block are $\mathcal{O}(HM)$, $\mathcal{O}(HZ + Z^2)$ and $\mathcal{O}(H^2)$, respectively. The work \cite{vaswani2017attention} summarizes the complexities of RNN and self-attention, which are $\mathcal{O}(M)$ with sequence length of $1$ and $\mathcal{O}(H^2)$ with feature length of $1$. Thus, the complexity of the NN model is $\mathcal{O}(HM+HZ+Z^2+2H^2+M)=\mathcal{O}(2.5HN+N^2/4+2H^2+N)$. 

Specifically, the adaptive filter length $N$ is fixed at $1024$, the subband number $K$ is $32$, the downsampling factor $D$ is $K/2=16$ and the subband filter length $Q$ is $N/D=64$. Therefore, $M$ is $2DQ=2048$, $H$ is $128$ and $Z$ is $DQ/2=512$. By using python package $thop$, the number of parameters of our model is $1119752$, and the number of floating point operations per second (FLOPS) is $1419520$, which means if we need to run the model at $16$ kHz without skips, the FLOPS a chip or a computer can perform need to be at least $1.42$ G. Thanks to the development of today's neural processing unit and hardware accelerating techniques, many machines can operate our model in real time. We use a computer with an Intel(R) Core(TM) i9-14900KF 3.20 GHz processor and apply the Open Neural Network Exchange (ONNX) \cite{onnxruntime} to our model to speed up the inference. The average, maximum, and median computation times at each iteration are measured as $0.17$ ms, $0.25$ ms, and $0.17$ ms, respectively. Recalling that $\frac{1}{f_sD}=1$ ms, the proposed model can operate in real time without skip updating. If the computing machine cannot operate the model without skips, the skip updating factor $B$ can be adjusted upward.

\section{Experimental Setup}
\label{SEC:ES}

\subsection{Performance Metric}
\label{SUB:1}
For evaluation metric, the normalized mean squared error (NMSE), which is the average ratio of the error signal power to the desired signal power is often used. The NMSE in dB is:
\begin{equation}
\mathrm{NMSE} = 10\mathrm{log}_{10}\frac{\sum_{i=0}^{T_t}E\{e^2(i)\}}{\sum_{i=0}^{T_t}E\{d^2(i)\}}
\label{EQ:NMSE}
\end{equation}
where $T_t$ is the number of sound samples. The NMSE is computed based on 50 independent runs. The value of NMSE is usually lower than $0$. The lower value of NMSE indicates better noise attenuation.

\subsection{Experimental Settings}

A large variety of noise sources are used during the training phase to train the proposed model. The training set includes a subset of the Environmental Sound Classification (ESC-50) dataset \cite{piczak2015esc} and the entire Nonspeech dataset \cite{hu2008segregation}. The ESC-50 dataset consists of 2000 recordings, including animal and urban sounds. The test set is also obtained from NOISEX-92 \cite{varga1993assessment}. All recordings in NOISEX-92 are recorded onto digital audio tape using a $1/2"$ $B\&K$ condenser microphone. The four subsets of NOISEX-92 are Speech Babble, Factory Floor Noise, Cockpit Noise and Engine Noise. Speech Babble corresponded $100$ people speaking in a canteen. Factory Floor Noise was recorded near the plate-cutting and electrical welding equipment. Cockpit Noise and Engine Noise were recordings for a moving buccaneer jet and in an engine room, respectively. The sampling rate of both the training and test sets is fixed at $16$ kHz.

To simulate the physical structure of the environment for the ANC system, a 3D rectangular enclosed room is often used \cite{kestell2000active, cheer2012active}. The room impulse response (RIR) is computed by image method \cite{allen1979image}. In this paper, a spatial arrangement similar to \cite{zhang2021deep} is used. The size of room is $5$ m $\times$ $4$ m $\times$ $3$ m (length $\times$ width $\times$ height), the secondary speaker is located at $(3, 2, 1.5)$ and the error microphone is located at $(3.5,2,1.5)$. In order to simulate the variations of the primary path, reference microphones are placed at $9$ positions. These positions are the vertices and center of a cube space centered at $(1, 2, 1.5)$ with edge length of $1$. Omnidirectional microphones are employed to record sound signals. The lengths of the primary path and secondary path are set as $2048$ and $1024$, respectively. The sound velocity is $340$ m/s. Noting that low revereration time is often used in deep ANC algorithms \cite{zhang2021deep, cha2023dnoisenet}, the reverberation time (T60) is employed as $0.15$ s. For training a meta learning model, different signal-to-noise ratios (SNRs) of measurements are selected as $0$, $5$, $10$, $15$, $20$, $25$, $30$ dB. The SNRs for testing algorithms are set as $5$, $15$ and $25$ dB. Notice that the saturation effect of the secondary speaker is the most significant nonlinearity in the ANC system \cite{tobias2002performance, tobias2006lms}. The nonlinearity factors $\eta$ in (\ref{EQ:SEF}) are set as $0.1$, $1$, $10$ and $\infty$ during the training phase, while in the test, they are $0.5$, $2$ and $\infty$. Every audio is clipped as $3$ s without changes of the primary path when training our model, while the audio length of the testing set is fixed at $10$ s, and there is a sudden change of the primary path at the middle of the ANC process. We combine the clipped audios and the simulated acoustic paths to generate $9000$ for training, where $90\%$ of the recordings are training set and the rest are for validation. The adaptive filter length is fixed at $1024$. The step size $\mu$ is set as $0.4$. Since a meta-learning-based training method is used in this paper, there is no labeled data and the measurement noise will affect the error noise. The number of meta frames $F$ is $8$. The batch size of training data is set at $150$ to accelerate the training phase, and we set an early stopping time to avoid overfitting with patience of $3$. Meanwhile, when training our model, the skip updating factor $B$ is set at $0$. 

\begin{table}
\setlength{\tabcolsep}{2pt}
\caption{$\mathrm{NMSE}$s at $\eta^2=0.5$ with different SNRs}
\label{TB:PF_1}
\scalebox{.8}{
\begin{tabular}{lllllllllllll}
\toprule
Noise type  & \multicolumn{3}{l}{Babble} & \multicolumn{3}{l}{Factory Floor} & \multicolumn{3}{l}{Cockpit} & \multicolumn{3}{l}{Engine}\\
\cline{2-4}\cline{5-7}\cline{8-10}\cline{11-13}
SNR (dB) & $5$ & $15$ & $25$ & $5$ & $15$ & $25$ & $5$ & $15$ & $25$ & $5$ & $15$ & $25$\\
\midrule
NFxLMS & $-5.98$ & $-6.83$ & $-6.77$ & $-5.35$ & $-6.17$ & $-6.61$ & $-1.85$ & $-1.94$ & $-1.98$ &  $-4.12$ & $-4.30$ & $-4.31$\\
DSNFxLMS & $-5.60$ & $-6.42$ & $-6.28$ & $-5.47$ & $-6.39$ & $-6.78$ & $-2.59$ & $-2.74$ & $-2.81$ & $-4.77$ & $-4.97$ & $-5.06$\\
MDSAF-MD & $-7.24$ & $-8.55$ & $-9.10$ & $-7.38$ & $-8.70$ & $-9.30$ & $-6.84$ & $-7.69$ & $-8.05$ & $-7.27$  & $-8.11$ & $-8.21$\\
MDSAF   & $\textbf{-7.50}$ & $\textbf{-8.97}$ & $\textbf{-9.54}$ & $\textbf{-7.85}$ & $\textbf{-9.29}$ & $\textbf{-10.17}$ & $\textbf{-7.13}$ & $\textbf{-8.05}$ & $\textbf{-8.48}$ & $\textbf{-7.47}$ & $\textbf{-8.42}$ & $\textbf{-8.55}$\\
\bottomrule
\end{tabular}
}
\end{table}

\begin{table}
\setlength{\tabcolsep}{2pt}
\caption{$\mathrm{NMSE}$s at $\mathrm{SNR}=5$ dB with different $\eta^2$}
\label{TB:PF_2}
\scalebox{.8}{
\begin{tabular}{llllllllllllllllllllllllll}
\toprule
Noise type  & \multicolumn{3}{l}{Babble} & \multicolumn{3}{l}{Factory Floor} & \multicolumn{3}{l}{Cockpit} & \multicolumn{3}{l}{Engine}\\
\cline{2-4}\cline{5-7}\cline{8-10}\cline{11-13}
$\eta^2$ & $0.5$ & $2$ & $\infty$ & $0.5$ & $2$ & $\infty$ & $0.5$ & $2$ & $\infty$ & $0.5$ & $2$ & $\infty$ \\
\midrule
NFxLMS & $-5.98$ & $-6.29$ & $-5.95$ & $-5.35$ & $-5.64$ & $-5.81$ & $-1.85$ & $-1.85$ & $-1.85$ &  $-4.12$ & $-4.05$ &  $-4.20$\\
DSNFxLMS & $-5.60$ & $-5.90$ & $-5.61$ & $-5.47$ & $-5.76$ & $-5.90$ & $-2.59$ & $-2.57$ & $-2.60$ &  $-4.77$ & $-4.70$ & $-4.80$\\
MDSAF-MD & $-7.24$ & $-7.49$ & $-7.09$ & $-7.38$ & $-7.45$ & $-7.42$ & $-6.84$ & $-6.83$ & $-6.87$ &  $-7.27$ & $-7.10$ & $-7.15$\\
MDSAF  & $\textbf{-7.50}$ & $\textbf{-7.74}$ & $\textbf{-7.42}$ & $\textbf{-7.85}$ & $\textbf{-7.93}$ & $\textbf{-8.03}$ & $\textbf{-7.13}$ & $\textbf{-7.14}$ & $\textbf{-7.17}$ & $\textbf{-7.47}$  & $\textbf{-7.22}$ & $\textbf{-7.39}$\\
\bottomrule
\end{tabular}
}
\end{table}

\subsection{Competing Algorithm}
To evaluate the learning-based adaptive filter performance, we need to select some competing algorithms. It is worth mentioning that despite supervised learning-based ANC algorithms \cite{zhang2021deep, zhang2022deep, zhang2023deep, luo2023deep} have gain significant success, we do not choose one of them. Supervised learning methods use true labels to essential information about the system to be modeled. In this paper, we try to perform data-driven techniques without true labels. Therefore, we still choose two traditional adaptive algorithms for performance evaluation. They are the NFxLMS and delayless subband normalized filtered-x least mean square (DSNFxLMS) algorithms for performance comparison. The updating rule of the NFxLMS algorithm is given as:
\begin{equation}
\textbf{w}(n+1)=\textbf{w}(n) - \mu \frac{\textbf{x}_f(n)e(n)}{||\textbf{x}_f(n)||^2+\epsilon} 
\label{EQ:NLMS}
\end{equation}
where $\epsilon$ is a small positive number to prevent the denominator term from $0$. It is worth mentioning that the simple normalization factor $||\textbf{x}_f(n)||^2$ is not practical. The equivalent delay of the secondary path \cite{ardekani2010theoretical} also affects the normalization factor. In this paper, the secondary path is fixed so that the equivalent delay of the secondary path is fixed. Thus, the step size of the NFxLMS algorithm can be adjusted to comprise the equivalent delay during the weight update. In addition, the updating rule of DSNFxLMS algorithm \cite{park2001delayless} is:
\begin{equation}
\textbf{w}_k(n+1)=\textbf{w}_k(n) - \mu \frac{\textbf{x}_{fk}(n)e_k(n)}{||\textbf{x}_{fk}(n)||^2+\epsilon} 
\label{EQ:DSNLMS}
\end{equation}

The nonstationary noise will affect the upper bound of the step size of adaptive algorithms. In addition, the saturation effect of the speaker function distorts the secondary source \cite{tobias2006lms}. To cope with all the testing scenarios, we fix the step sizes of NFxLMS and DSNFxLMS algorithms as $0.01$.

\section{Results and Discussion}
\label{SEC:ERC}
\subsection{Comparative Studies}
We first evaluate the performance of the proposed meta-learning-based delayless subband model with the competitors. The models with different secondary paths and the competitors are tested under three types of untrained noises in two nonlinear systems ($\eta^2=2$, $\eta^2=0.5$) and a linear system ($\eta^2=\infty$). Tables \ref{TB:PF_1} and \ref{TB:PF_2} present the averaged NMSEs under different noise types with various SNR levels and nonlinearity factors. Traditional algorithms give larger NMSEs when using the same step size across all test scenarios. This is because, even in the linear condition, the acoustic path length is relatively long, and the noise sources are not stationary. The MDSAF and MDSAF-MD demonstrate superior performance across all cases.

\begin{figure*}[htb]
\centering
\subfigbottomskip=2pt 
\subfigcapskip=2pt
\subfigure[Babble]{
\includegraphics[scale=0.8]{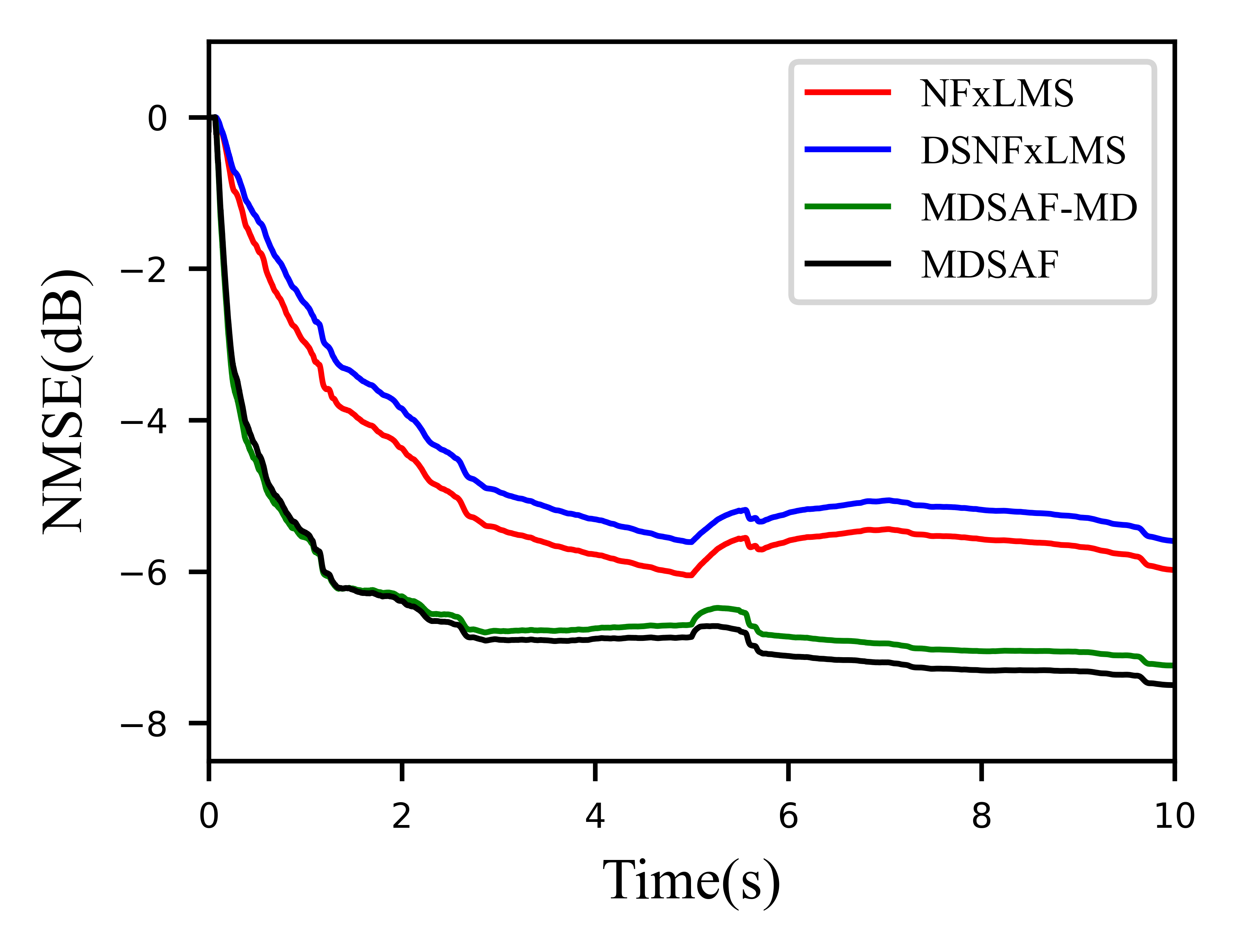}}
\hspace{-0.2cm}
\subfigure[Factory Floor]{
\includegraphics[scale=0.8]{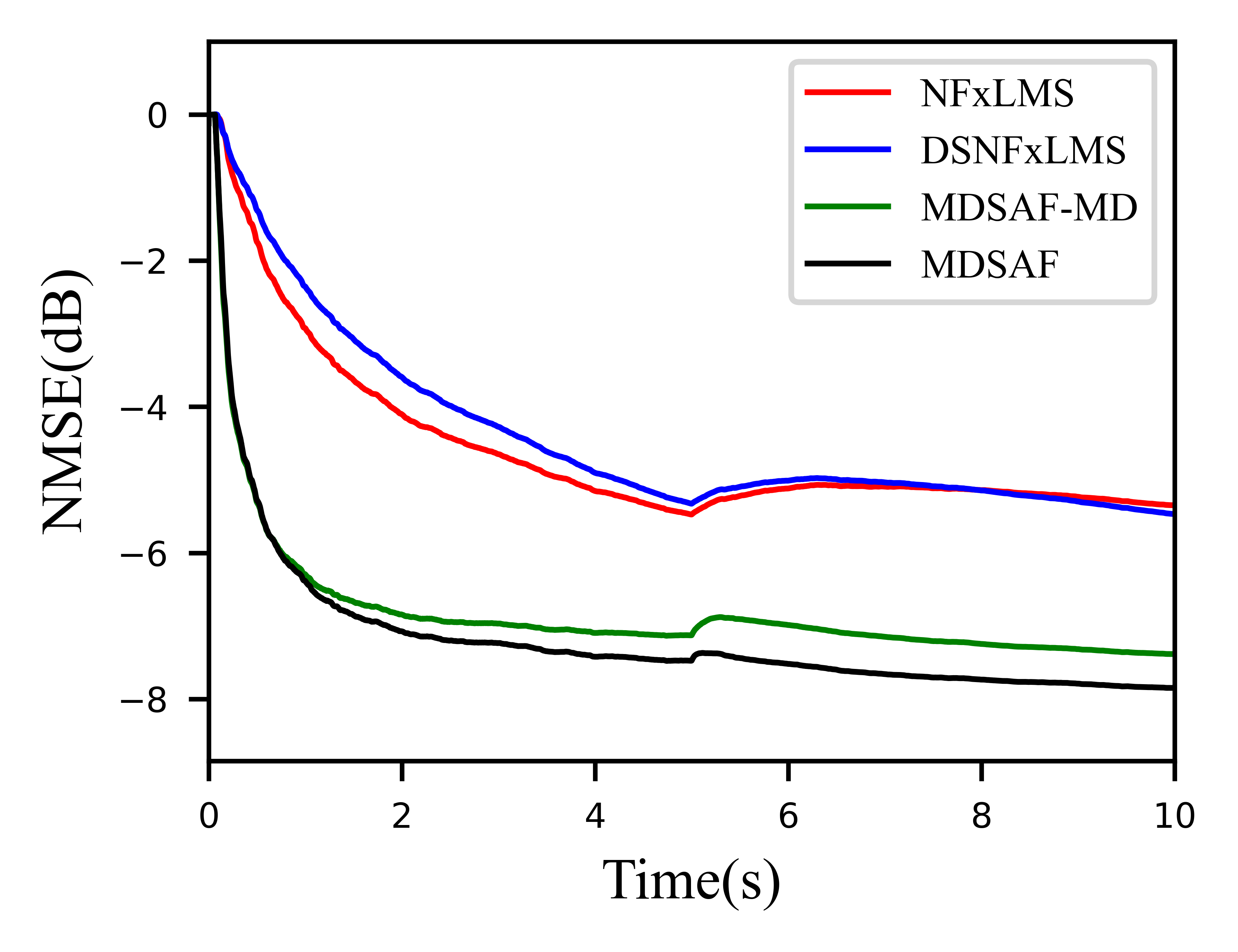}}
\\
\vspace{0cm}
\subfigure[Cockpit]{
\includegraphics[scale=0.8]{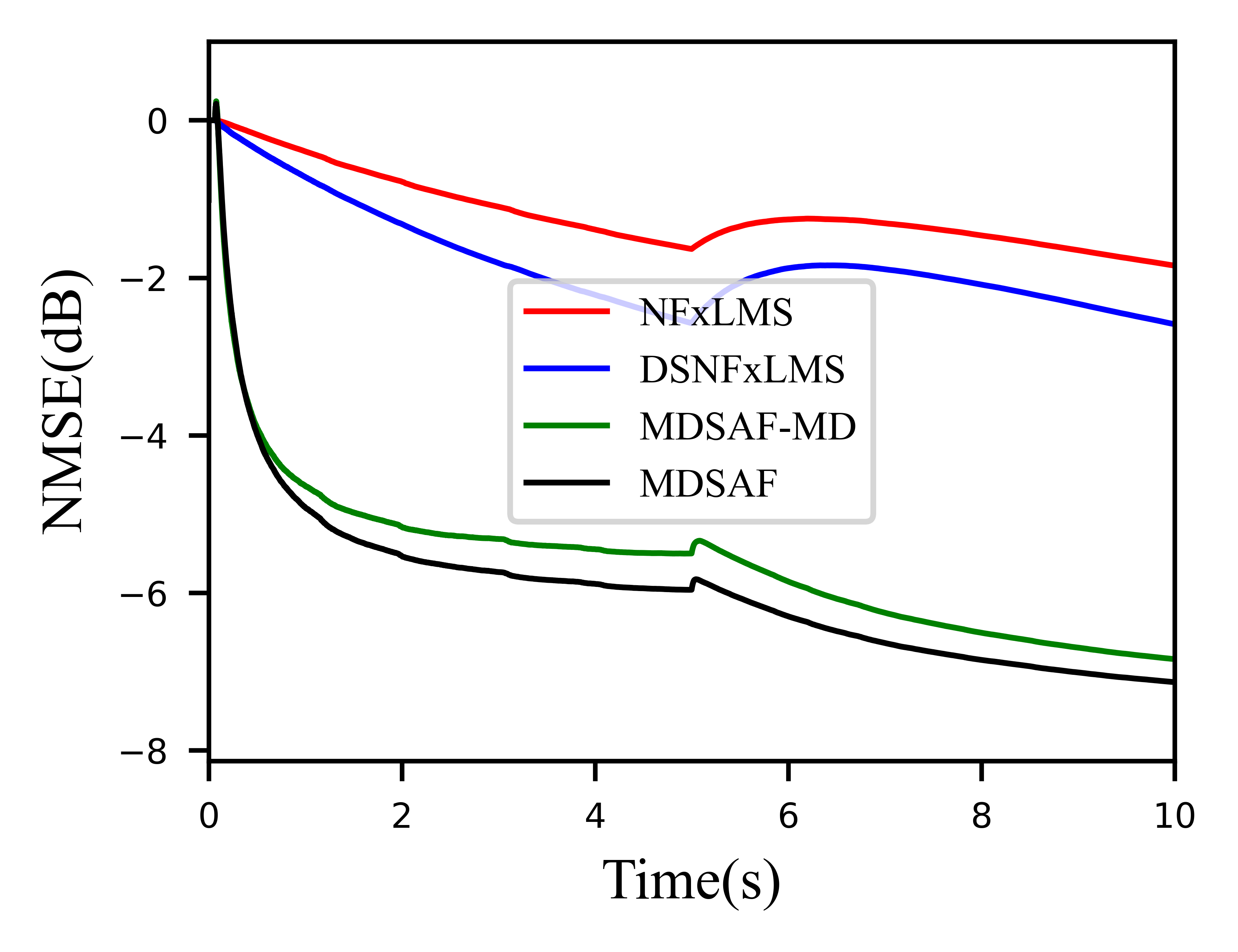}}
\hspace{-0.2cm}
\subfigure[Engine]{
\includegraphics[scale=0.8]{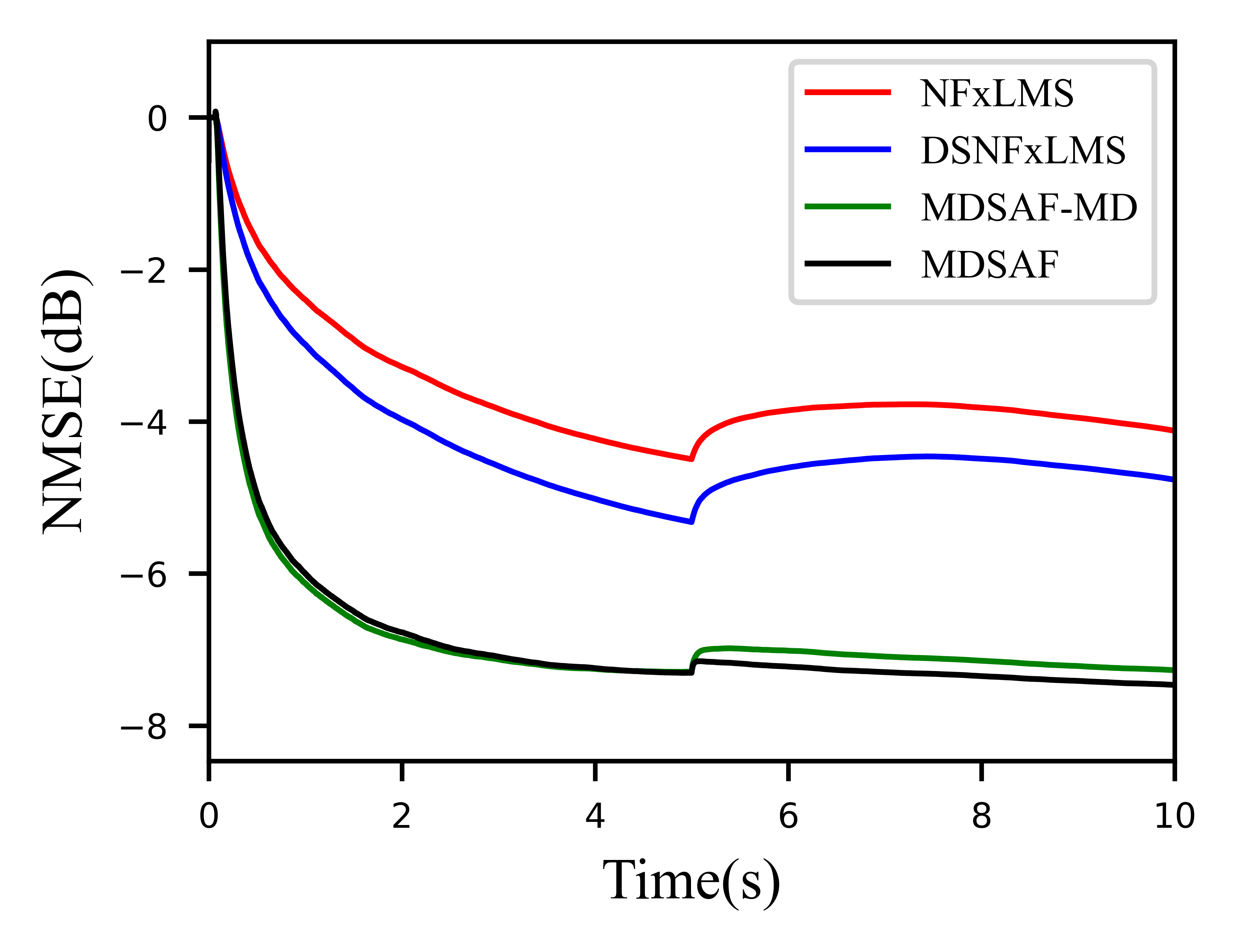}}
\caption{$\mathrm{NMSE}$s under different noise types at SNR$=5$ dB and $\eta^2=0.5$}
\label{FIG:NONLINEAR_NMSE}
\end{figure*}

\begin{figure*}[htb]
\centering
\subfigbottomskip=2pt 
\subfigcapskip=2pt
\subfigure[Babble]{
\includegraphics[scale=0.8]{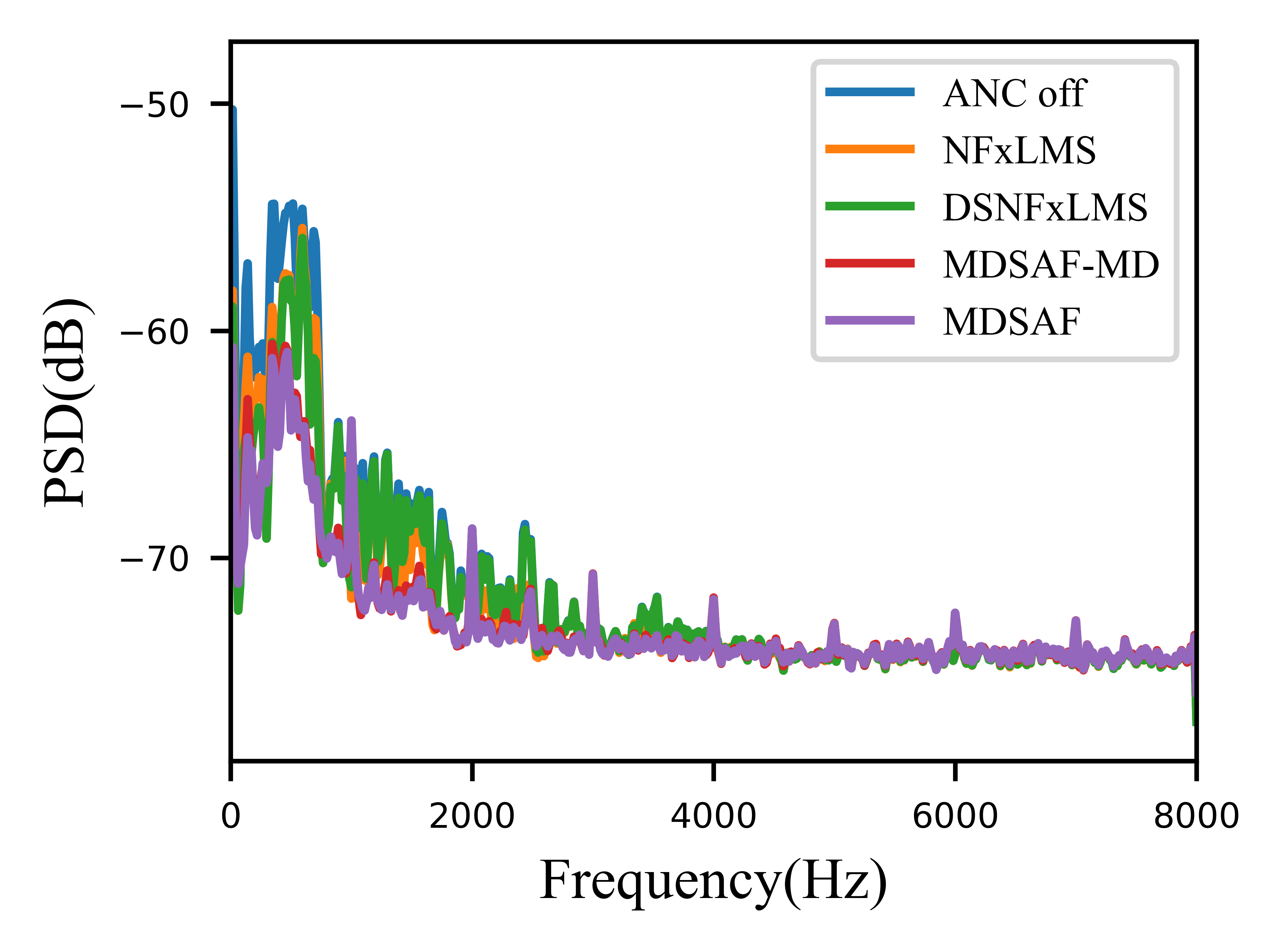}}
\hspace{-0.2cm}
\subfigure[Factory Floor]{
\includegraphics[scale=0.8]{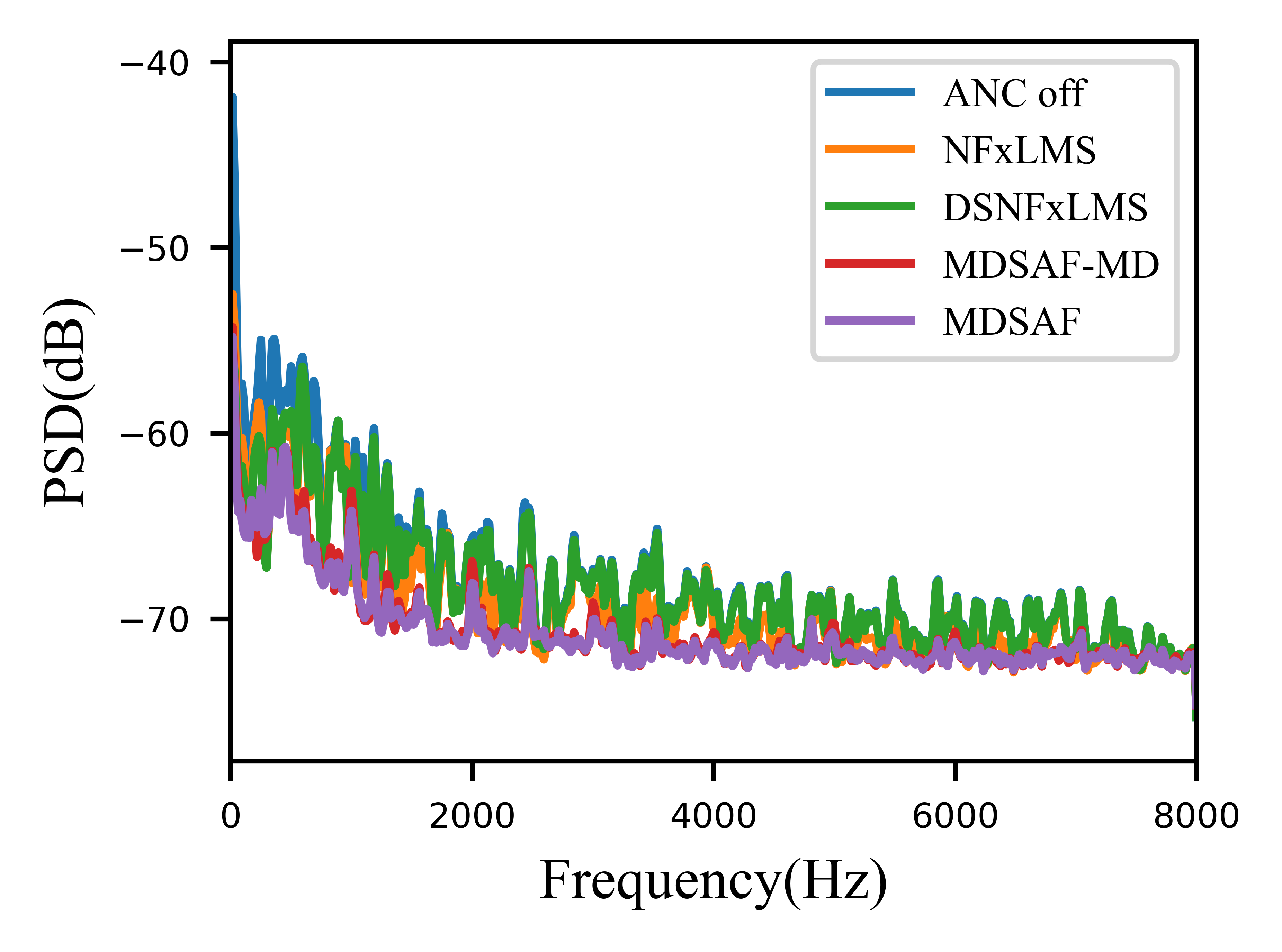}}
\\
\vspace{0cm}
\subfigure[Cockpit]{
\includegraphics[scale=0.8]{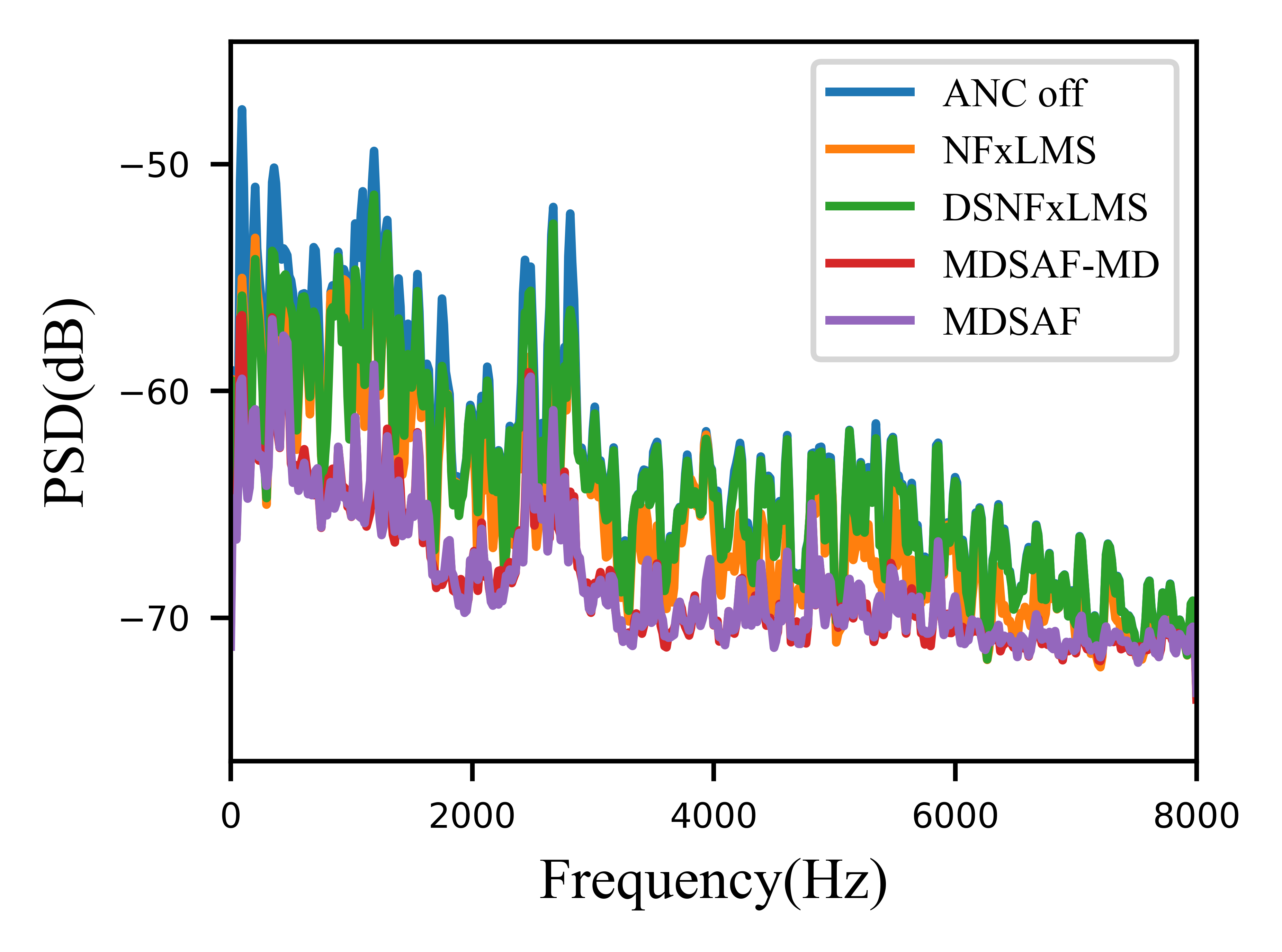}}
\hspace{-0.2cm}
\subfigure[Engine]{
\includegraphics[scale=0.8]{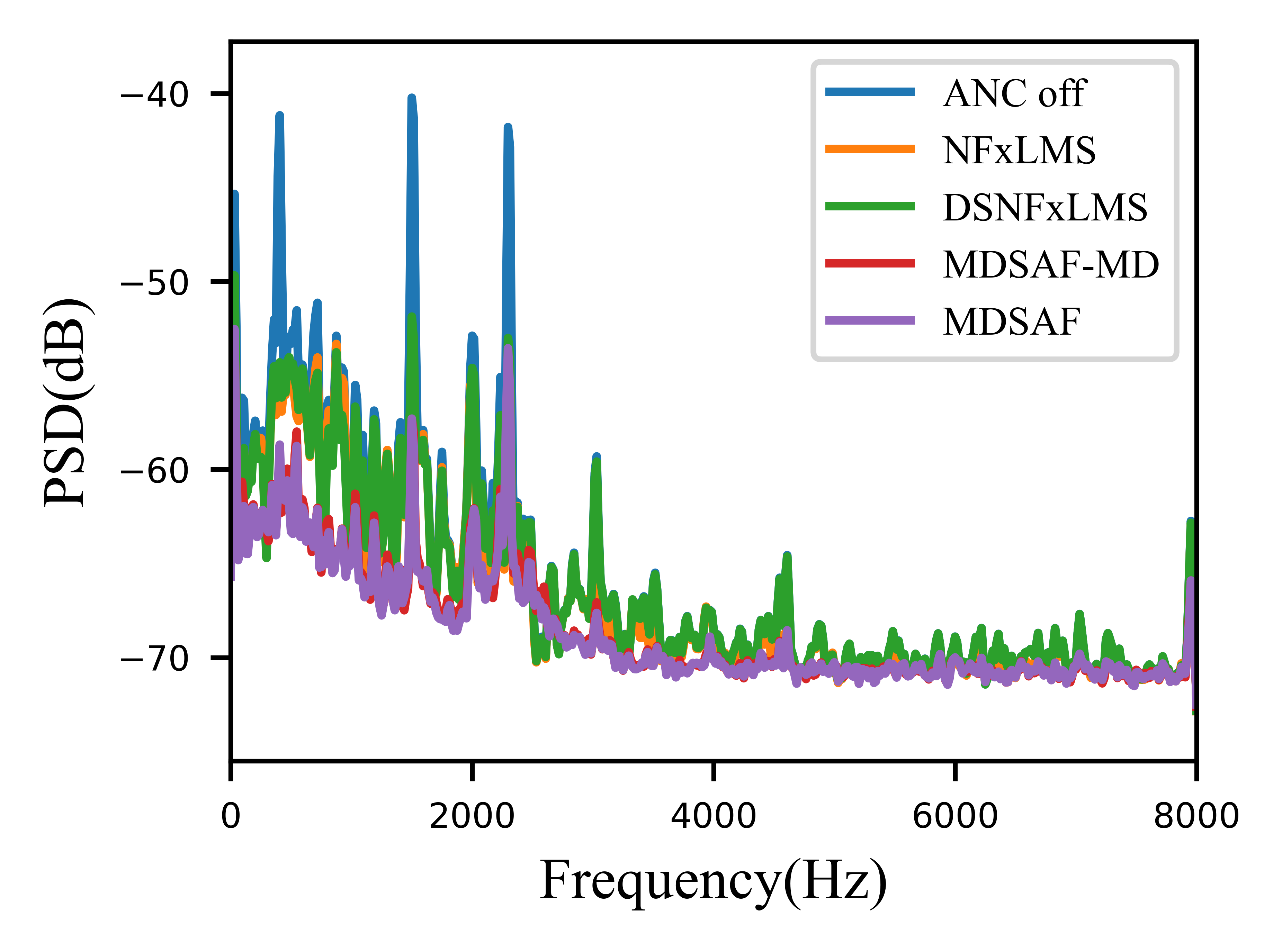}}
\caption{PSDs under different noise types at SNR$=5$ dB and $\eta^2=0.5$}
\label{FIG:NONLINEAR_PSD}
\end{figure*}

Figs. \ref{FIG:NONLINEAR_NMSE} and \ref{FIG:NONLINEAR_PSD} display the NMSE and power spectrum curves for further comparison. The power spectrum, which measures relative signal power in the frequency domain, illustrates the noise attenuation achieved at different frequencies. The results in Fig. \ref{FIG:NONLINEAR_NMSE} are obtained with four types of testing noise and a nonlinearity factor of $\eta^2=0.5$. The MDSAF and MDSAF-MD achieve lower NMSEs upon convergence than others and demonstrate faster convergence during the ANC process. Interestingly, even without training the MDSAF and MDSAF-MD for primary path changes, the MDSAF and MDSAF-MD adapt well and converge faster than competitors. Notably, the MDSAF and MDSAF-MD are trained using only noisy data, which means they never see the true desired signal, yet they effectively compensate for nonlinearity.

In Fig. \ref{FIG:NONLINEAR_PSD}, the MDSAF and MDSAF-MD achieve wideband noise reduction, whereas the competing methods are effective only at low frequencies. This aligns with the findings of \cite{samarasinghe2016recent}, highlighting the limitations of conventional ANC systems at low and medium frequencies. In Fig. \ref{FIG:NONLINEAR_PSD}(a), the main frequency band of the noise is below $2000$ Hz. While the MDSAF and MDSAF-MD perform well in the low-frequency band, some power increases exist in some high-frequency bins. This is because the delayless subband filters have some stacking errors, and meta-learning-based models have limited prediction performance in low-power frequency bands. It is important to note that we have used the same architecture to train two models under different secondary path conditions. The model with the whole secondary path handles a more straightforward ANC problem, so a decrease in performance is expected when using the model trained with the main delay of the secondary path. Notably, the model with the whole secondary path can most effectively cancel unwanted noise in both linear and nonlinear situations.

\section{Conclusion}
\label{SEC:CC}
This paper takes inspiration from deep ANC and meta-adaptive filter approaches, transferring the main idea of a meta-learning-based model into a feedforward ANC system. We reformulate the ANC problem as a meta-learning problem and employ a NN model as the updating rule of the adaptive filter. Our architecture differs from the original meta-adaptive filter since we modify the delayless subband architecture to avoid multi-point updating of the adaptive filter weight in the frequency domain. A modified single-headed recurrent neural network is used. We design a learnable feature embedding method to learn the temporal feature across the position of subband signals. Furthermore, using the delayless subband architecture and the skip updating strategy, our learning-based architecture can operate in real time without violating the causality constraint. Furthermore, we apply the proposed architecture to an ANC system that does not use the whole secondary path. We experimented with our architecture on various paths and different types of noise. The results show that with the proper training data and loss function, our architecture can cancel noise under path and noise changes without true labels, which is often required in other deep learning-based ANC algorithms. Moreover, the proposed model can cope with situations not only with the whole secondary path but also with the main delay of the secondary path. Extensive comparative and parametric studies illustrate our algorithms' superior generalization and performance.

\bibliographystyle{elsarticle-num} 
\bibliography{refs}


\end{document}